\documentclass[sn-basic]{sn-jnl}

\usepackage{graphicx}%
\usepackage{multirow}%
\usepackage{amsmath,amssymb,amsfonts}%
\usepackage{amsthm}%
\usepackage{mathrsfs}%
\usepackage[title]{appendix}%
\usepackage{xcolor}%
\usepackage{textcomp}%
\usepackage{manyfoot}%
\usepackage{booktabs}%
\usepackage{algorithm}%
\usepackage{algorithmicx}%
\usepackage{algpseudocode}%
\usepackage{listings}%
\usepackage{colors}%
\usepackage{orcidlink}%

\theoremstyle{thmstyleone}%
\theoremstyle{thmstyletwo}%

\theoremstyle{thmstylethree}%

\begin{document}

\title[Deciphering the Origin of the Elements]{Deciphering the Origins of the Elements Through Galactic Archeology}


\author[1]{\fnm{Khalil} \sur{Farouqi}}\email{kfarouqi@lsw.uni-heidelberg.de}

\author[2,3]{\fnm{Anna} \sur{Frebel}}\email{afrebel@mit.edu, ORCID: 0000-0002-2139-7145}

\author*[4,5]{\fnm{Friedrich-Karl} \sur{Thielemann}}\email{f-k.thielemann@unibas.ch, ORCID: 0000-0002-7256-9330}

\affil[1]{\orgdiv{ZAH, Landessternwarte}, \orgname{University of Heidelberg}, \orgaddress{\street{K\"onigstuhl 12}, \city{69117 Heidelberg}, \country{Germany}}}

\affil[2]{\orgdiv{Department of Physics}, \orgname{Massachusetts Institute of Technology}, \city{Cambridge, MA 02139}, \country{USA}}

\affil[3]{\orgdiv{Kavli Institute for Astrophysics and Space Research}, \orgname{ Massachusetts Institute of Technology} \city{Cambridge, MA 02139}, \country{USA}}
 
\affil[4]{\orgdiv{Physics Department}, \orgname{University of Basel}, \orgaddress{\street{Klingelbergstrasse 82}, \city{4056 Basel}, \country{Switzerland}}}

\affil[5]{\orgdiv{Theory}, \orgname{GSI Helmholtz Center for Heavy Ion Research}, \orgaddress{\street{Planckstrasse 1}, \city{64291 Darmstadt}, \country{Germany}}}


\abstract{Low-metallicity stars preserve the signatures of the first stellar nucleosynthesis events in the Galaxy, as their surface abundances reflect the composition of the interstellar medium from the time when they were born. Aside from primordial Big Bang nucleosynthesis, massive stars, due to their short lifetimes, dominate the wind and explosive ejecta into the interstellar medium of the early Galaxy. Most of them will end as core-collapse supernova (CCSN) explosions, and typical ejected abundance distributions, e.g. in terms of the $\alpha$-element-to-Fe ratios, reflect these contributions. Essentially all CCSNe contribute $^{56}$Fe (decaying from radioactive $^{56}$Ni). Therefore, low-metallicity stars can be used to test whether the abundances of any other elements are correlated with those of Fe, i.e. whether these elements have been co-produced in the progenitor sources or if they require either a different or additional astrophysical origin(s).
The present analysis focuses on stars with [Fe/H]$<$-2, as they probe the earliest formation phase of the Galaxy when only one or very few nucleosynthesis events had contributed their ejecta to the gas from which the lowest metallicity stars form. This was also the era before low and intermediate mass stars (or type Ia supernovae) could contribute any additional heavy elements.
Following earlier works into the origin of heavy r-process elements \citep{Farouqi.Thielemann.ea:2022}, we extend the present study to examine Pearson and Spearman correlations of Fe with Li, Be, C, N, Na, Mg, Si, S, Ca, Ti, Cr, Ni, Zn, Ge, Se, Sr, Zr, Ba, Ce, Sm, Eu, Yb, Lu, Hf, Os, Ir, Pb, Th, and U, using high-resolution stellar abundance data from the SAGA \citep{Sagadatabase} and JINA \citep{JINAbase:2018} databases. The main goal is to identify which of the observed elements (i) may have been co-produced with Fe in (possibly a variety of) CCSNe, and which elements require (ii) either a completely different, or (iii) at least an additional astrophysical origin.
}




\maketitle

\section{Introduction}\label{sec1}


Stellar spectra reflect the chemical abundances of the photosphere. With understood exceptions, this surface composition well represents the star's birth chemical composition because internal burning stages do not alter the outer layers in relatively unevolved low-mass stars, including those at low metallicity. Therefore, any star preserves the original composition of its pre-stellar cloud from which the star was born. Very and Extremely Metal-poor (VMP/EMP) stars (with [Fe/H] $<-$2.0 and [Fe/H] $<-$3.0; respectively) offer invaluable insights into the early stages of galactic evolution and nucleosynthesis, because at metallicities of [Fe/H]$<-$2.5 only one (or very few!) progenitor sources plausibly contributed to the enrichment of the local gas. Thus, stellar abundance observation in the VMP/EMP stars allow us to study the ejecta composition of individual nucleosynthesis sites that occurred in the early universe \citep[see e.g.][]{Nomoto.Tominaga.ea:2006,Honda.Aoki.ea:2007,Frebel:2018,Yong.ea:2021,Farouqi.Thielemann.ea:2022,Placco.ea:2023,Frebel:2024}.

As part of the chemical evolution of the Galaxy, intermediate and low-mass (AGB) stars are thought to start to contribute heavy s-process elements with a time delay. Such corresponding signatures are found only at higher metallicities from [Fe/H] $>-$2 to $-$1.5 \citep[e.g.][]{sneden08}. Type Ia supernovae, originating from white dwarf explosions, are further delayed and only appear since the chemical evolution has reached the level of [Fe/H] = $-$1. As a result, the metallicity range of VMP/EMP stars probes  the earliest phases of galactic evolution when essentially only fast evolving massive stars contributed via their stellar winds or core-collapse explosions, with typical yields of about $0.1$M$_\odot$ of Fe per CCSN \citep[see e.g.][]{matteucci86,Timmes.Woosley.Weaver:1995,Kobayashi.Umeda.Nomoto.ea:2006,Nomoto.ea:2013,Kobayashi.Karakas.Lugaro:2020}. 
Consequently, only VMP/EMP stars can reveal the clean signatures of specific nucleosynthesis sources and sites in their spectra and through analysis of their observed elemental abundance patterns. On the contrary, higher metallicity stars formed from gas enriched by many sources and their abundance patterns reflect averaged values from many contributing sites and events.

Detailed abundance patterns have been determined from hundreds of low-metallicity stars over the last 30 years. Yet, the interpretation of these signatures remains ongoing as our global understanding of the astrophysical production of all elements across the Periodic Table continues to expand and evolve. 
In this regard, a Pearson correlation coefficient of 1 for element X with respect to the Fe abundance points to a linear relation of the abundances of element X as Fe increases. At low metallicities, where no other Fe producing sites are available, such a correlation points to a CCSN origin for element X. Applying the same procedure to stellar explosive ejecta that are mixed into a star-forming cloud by one type of contributing event (e.g., a CCSN) will always lead to an X/Y ratio typical for this event. This is independent of the total amount of matter introduced by a nearby or not-so-nearby source of the same type. While the total values of X and Y could vary between star forming clouds, the ratio X/Y would be constant and an increase in Y would be accompanied by a linear increase in X. If there exists a significant correlation with Fe, but with values smaller than 1, a partial CCSN origin is indicated, but an additional, independent source is required. In case of a vanishing correlation (close to 0), this element must originate from another site completely different than CCSNe. Further statistical tools for these kinds of analyses are discussed in the Appendix, in addition to the ones presented in a previous publication \citep{Farouqi.Thielemann.ea:2022}. 

We have applied these statistical tools to the entire content of the “Stellar Abundances for Galactic Archaeology” SAGA database \citep{Sagadatabase} and JINAbase \citep{JINAbase:2018} that report  element abundance observations in stars for metallicities of [Fe/H] $<-$2 from across the literature. The goal is to assess the body of data for any underlying correlations among various elements in the Periodic Table to gain further insight into their cosmic origins and production pathways. 
We combine new results with respect to the statistical analysis carried out with a review-like pass through the possible contributing nucleosynthesis sites and will therefore also display figures and results from previously published related articles, in order to further manifest the connection between stellar abundance observations and stellar model and nucleosynthesis predictions.

This paper starts with a detailed correlation analysis and provides the results in Section \ref{analysis}. The sections following this analysis will discuss all elements where observational data are available, interpreting the results of the correlation analysis in terms of astrophysical sources, before presenting a summary and conclusions.

\section{Correlation Analysis and Significance Tests}
\label{analysis}
In this Section, we present the results of our correlation analysis, examining the relationships between Fe and various elements within our dataset. We calculate as the main result of the analysis the Pearson correlation coefficients for Fe and the elements from Li to Th and U.

The Pearson correlation coefficient quantifies the strength and direction of the linear relationship between two variables, in this case, iron and each of the elements considered here. A PCC = 1 indicates a perfect positive linear relationship, while values close to 0 stand for the complete absence of a linear relationship (for further details, see the appendices of \cite{Farouqi.Thielemann.ea:2022} and this study).
Results for the PCC values are presented in column 2 of Table~\ref{tab:correlation_resultsB}.

Other important entries in this table are the p-values (second column of Table~\ref{tab:correlation_resultsB})
to assess the statistical significance of the correlations. A small p-value suggests that the observed correlation is unlikely to occur due to random chance alone. Typically, a predetermined significance level, such as $<$0.05 or $<$0.01, is chosen for the correlation test (third column), we decided for 0.01. If the p-value is less than the significance level, we conclude that there is a statistically significant correlation between the variables. Otherwise, if the p-value is greater than the significance level, it indicates no significant correlation. Therefore, we consider also the p-value (we chose p = 0.01 as threshold, but 0.05 still provides strong evidence), in addition to the calculated Pearson correlation coefficients, in order to determine the significance of the correlation. 

For further detailed discussions of statistical tools, going beyond the Pearson and Spearman correlations, as well as p-values, an extended overview of the literature is given in \cite{tamhane00} and \cite{spiegelhalter19}. However, for context, we briefly address here the remaining entries in Table~\ref{tab:correlation_resultsB}. An important quantity is the number of available data points (last column), i.e. the number of stars where Fe and the element of interest are observed. To  reliably measure the correlation of two quantities, typically the minimum number of required data points is around 20 to 30. Mutual information (MI; see column 5) measures whether a dependence between both entries (in our case the Fe and other element abundance) is existing. For MI values very close to 0 this is not the case, so we chose a limiting value of 0.02 to indicate at least a weak dependence, and 0.3 for indicating a strong dependence (shown in column 6). 

\begin{table}
  \centering
  \caption{PCC values of Fe with elements from Li to U for $\mbox{[Fe/H]}<-2$; significance and other column headings, see the text/appendices; CEMP and NEMP stars defined as [C/Fe] $\ge$ 0.7, [N/Fe]$\ge$ 0.5; for O- and Na-enhanced metal-poor stars (OEMP and NaEMP) we use [O/Fe]$\ge$ 1 and [Na/Fe]$\ge$ 1;
  correlation colors: blue (no), red (small), teal (moderate), green (strong), black (uncertain)}
  \label{tab:correlation_resultsB}
  \begin{tabular}{lllllll}
  \toprule
    Element Pair & PCC & P-Value & Significance &Mutual Info.  & Dependance & \# of Stars \\
    {\color{blue}Fe-Li}           & 0.08 & 0.05 &  no  & 0.009 & no  &  607 \\
    {\color{teal}Fe-Be}           & 0.61 & 0    & yes  & 0.23  & strong &   64 \\
    Fe-C (All)                    & 0.08 & 0    & yes  & 0.03  & weak & 1825 \\
    {\color{red}Fe-C} (CEMP)      & 0.25 & 0    & yes  & 0.1   & weak & 459 \\
    {\color{green}Fe-C} (Regular) & 0.70 & 0    & yes  & 0.34  & strong & 1366 \\
    Fe-N (All)                    & 0.22 & 0    & yes  & 0.07  & weak &  291 \\
    {\color{red}Fe-N} (NEMP)      & 0.27 & 0    & yes  & 0.1   & weak &  214 \\
    {\color{green}Fe-N} (Regular) & 0.85 & 0    & yes  & 0.36  & strong &  77 \\
    Fe-O (All)                    & 0.38 & 0    & yes  & 0.13  & weak &  338 \\
    {\color{red}Fe-O} (OEMP)      & 0.49 & 0    & yes  & 0.16  & weak &  74 \\
    {\color{green}Fe-O} (Regular) & 0.81 & 0    & yes  & 0.49  & strong &  264 \\
    {\color{red}Fe-Na}            & 0.06 & 0    & yes  & 0.06  & weak & 1189 \\
    {\color{green}Fe-Mg}          & 0.85 & 0    & yes  & 0.68  & strong & 1994\\
    {\color{green}Fe-Si}          & 0.79 & 0    & yes  & 0.53  & strong &  981\\
    {\color{green}Fe-S}           & 0.92 & 0    & yes  & 0.4   & strong &  85\\
    {\color{green}Fe-K}           & 0.81 & 0    & yes  & 0.41  & strong & 250 \\
    {\color{green}Fe-Ca}          & 0.89 & 0    & yes  & 0.73  & strong & 1964\\
    {\color{green}Fe-Ti}          & 0.82 & 0    & yes  & 0.5   & Strong & 1876\\
    {\color{green}Fe-Cr}          & 0.81 & 0    & yes  & 0.52  & Strong & 1620\\
    {\color{green}Fe-Ni}          & 0.92 & 0    & yes  & 0.85  & Strong & 1483\\
    {\color{green}Fe-Zn}          & 0.80 & 0    & yes  & 0.60  &strong &  733\\
    {\color{green}Fe-Ge}*         & 0.70 & 0.016 & no  & 0.30  &strong &  11\\
    {\color{green}Fe-Se}*         & 0.92 & 0.025 & no  & 0.77  &strong &  5\\
    {\color{red}Fe-Sr}            & 0.22 & 0    & yes  & 0.08  & weak & 1709\\
    {\color{red}Fe-Y}             & 0.16 & 0    & yes  & 0.08  & weak &  954\\
    {\color{red}Fe-Zr}            & 0.21 & 0    & yes  & 0.04  & weak &  569\\
    {\color{red}Fe-Mo}            & 0.17 & 0.27 & no   & 0.05  & weak & 43 \\
    {Fe-Ru}*            & 0.10 & 0.62 & no   & 0.04  & weak & 35 \\
    {Fe-Rh}*            & 0.75 & 0.052 & no   & 0.32  & strong & 7 \\
    {Fe-Pd}*           & 0.47 & 0.001 & yes   & 0.1  & weak & 43 \\
    {Fe-Ag}*            & 0.62 & 0.005    & yes  & 0.13  & weak & 18\\
    {\color{red}Fe-Ba}            & 0.10 & 0    & yes  & 0.03  & weak & 1701\\
    {\color{red}Fe-La}            & 0.13 & 0    & yes  & 0.15   & weak & 424\\
    {\color{red}Fe-Ce}            & 0.13 & 0.001& yes  & 0.15  & weak & 320\\
    Fe-Pr*            & 0.03 & 0.63 & no   & 0.06  & weak & 221\\
    Fe-Nd*                         & 0.03 & 0.57 & no  & 0.07  & weak & 414 \\
    {\color{red}Fe-Sm}            & 0.13 & 0.06 & no  & 0.06  & weak & 206\\
    {\color{red}Fe-Eu} (All)           & 0.16 & 0    & yes & 0.05  & weak & 695\\
    {\color{green}Fe-Eu} (r-limited) & 0.88 & 0    & yes & 0.66  & strong & 235\\
    {\color{red}Fe-Eu} (r-rich)    & 0.15 & 0    & yes & 0.05  & weak & 460\\
    {\color{red}Fe-Gd}            & 0.04 & 0.72 & no  & 0.05  & weak & 107\\
    {\color{red}Fe-Dy}            & 0.16 & 0    & yes & 0.04 & weak & 201  \\
    {\color{red}Fe-Ho}*            & 0.28 & 0.08 & no & 0.05  & weak   & 40  \\
    {\color{red}Fe-Er}*                         & 0.13 & 0.1  & no & 0.07  & weak  & 117 \\
    {\color{blue}Fe-Tm}*                         & 0.03 & 0.19  & no  & 0.04 & weak  & 48 \\
    {\color{red}Fe-Yb}            & 0.15 & 0.001 & yes & 0.1 & weak  & 154\\
    Fe-Lu*                         & 0.23 & 0.51  & no  & 0.27 & strong  & 10\\
    Fe-Hf*                         & 0.13 & 0.36  & no  & 0.07 & weak  & 50\\
    {\color{blue}Fe-Os}                         & 0.02 & 0.08  & no  & 0.01 &  no  & 58\\
    {\color{blue}Fe-Ir}           & 0.03 & 0.91  & no &  0.009  & no     & 20\\
    {\color{blue}Fe-Pb}           & 0.03 & 0.76  & no & 0.01  &  no      &   83 \\
    {\color{blue}Fe-Th}           & 0.06 & 0.67  & no & 0.01  &  no        & 48\\
    \botrule
  \end{tabular}
\end{table}

The results of the calculated Pearson correlation coefficient provide valuable information on the connections between the abundance of Fe and the abundances of a variety of other chemical elements considered here. 
Table \ref{tab:correlation_resultsB} presents a comprehensive overview of these correlations, highlighting the strength of the linear relationship between Fe and each element. The entries in the table are color-coded according to the level of correlation:
Blue: no correlation with Fe; 
Red: small correlation with Fe; 
Teal: moderately strong correlation with Fe; 
Green: strong correlation with Fe; and
Black: uncertain cases.

We now provide additional comments regarding the interpretation of the results. In addition to the colors, which identify the strength of the correlations, we also included additional labels (asterisks) to explain the significance of the results. As noted above, p-values smaller than 0.05 or 0.01  indicate the significance of the obtained result. We keep the color coding for values up to 0.1, but apply an asterisk to the PCC values for p-values greater than 0.1. The PCC value should then be taken as uncertain and we list it in black. The same is adopted for small data samples with less than about 40 entries. 

An exception from these categories are elements where no dependence with respect to the Fe abundance is found (MI in column 5 is less than 0.02), leading to a "no dependence" entry in column 6. As the p-value measures the significance of a correlation, a high value in column 3 underscores that there exists no correlation in such cases. This is also supported by the very small PCC values of less than 0.06. An additional exception includes the presentation of the PCC values for C, N, and O, when taken for the full data sample. The color coding is applied when utilizing the subsets of CEMP and "regular" stars (i.e., "all stars minus CEMP stars").

Upon initial inspection, a strong correlation indicates a clear origin of these elements as being co-produced with Fe in regular core-collapse supernovae (in green). There appears to be a strong to moderate contribution from regular CCSNe along with other contributing sources for elements shown in teal. A clear, but not dominant, contribution from CCSNe, likely requiring more prominent contributions from other sources, is represented by elements in red. Finally, elements which are apparently produced in other astrophysical sites than core-collapse supernovae are listed in blue.

More specifically, for the main category of elements from C to Yb, there exists a statistically significant correlation with Fe (indicated by small p-values). Within this category, there are three distinct groups. The first group exhibits a weak correlation with Fe, consisting of elements ranging from C to Na, if considering all stars (i.e. not excluding CEMP stars, but we also enter the related values for CEMP stars alone, as well as the remaining regular stars). The same holds for the elements from Sr to Yb. Be shows an intermediate behavior with a moderately strong correlation. Finally, the third group showcases a strong correlation with Fe, consisting of elements from Mg to Ge and Se, plus also C to Na, when considering only "regular" EMP stars, i.e. excluding CEMP stars.

For the remaining elements of the table, there is no statistically significant correlation with Fe, namely Li and elements beyond Yb (Z=70) or Os(Z=76). Hence, an independent origin (not related to CCSNe) is required.

In the following, we discuss all elements in an attempt to understand and interpret the correlations in terms of the nucleosynthesis sites, which are responsible for their production.

\section{The correlation of Fe with Li and Be}
For elements produced during standard Big Bang nucleosynthesis, i.e. Li \citep[e.g.][]{Arcones.Thielemann:2023}, or without a major stellar nucleosynthesis pathway, i.e., Be, no strong correlation with Fe is expected.

For Li, this behavior is supported by a very low Pearson correlation (although with reduced significance). While there exist other sources for Li, originating from low/intermediate mass stars and novae, those do not contribute to the chemical inventory at low metallicities [Fe/H]$<$-2 as considered in this study \citep[e.g.][]{Romano.ea:2021}.

However, the moderate correlation between Fe and Be is a bit surprising, suggesting that there appears to exist a relevant supernova contribution. The massive first stars did not burn hydrogen via the CNO cycle because of the absence of elements beyond Li. However, they were burning hydrogen at higher temperatures via hot pp-chains \citep{Wiescher.ea:1989,Wiescher.ea:2021}, where $^3$He is burned via an ($\alpha$,p)-reaction to $^7$Be. (There exist even speculations that also CNO nuclei could be produced this way, however, to a much smaller extent than in the later He-burning phase via the triple-$\alpha$ reaction). This channel of $^7$Be production via the hot pp chains, that leads to an ejection of Be from hydrostatic layers along with Fe during the later supernova explosion, appears to correlate the production of both elements at very low metallicities.

\section{The correlation of Fe with the $\alpha$-elements Mg to Ti}

\begin{figure}[h!]
    \centering
    \includegraphics[width=0.8\linewidth]{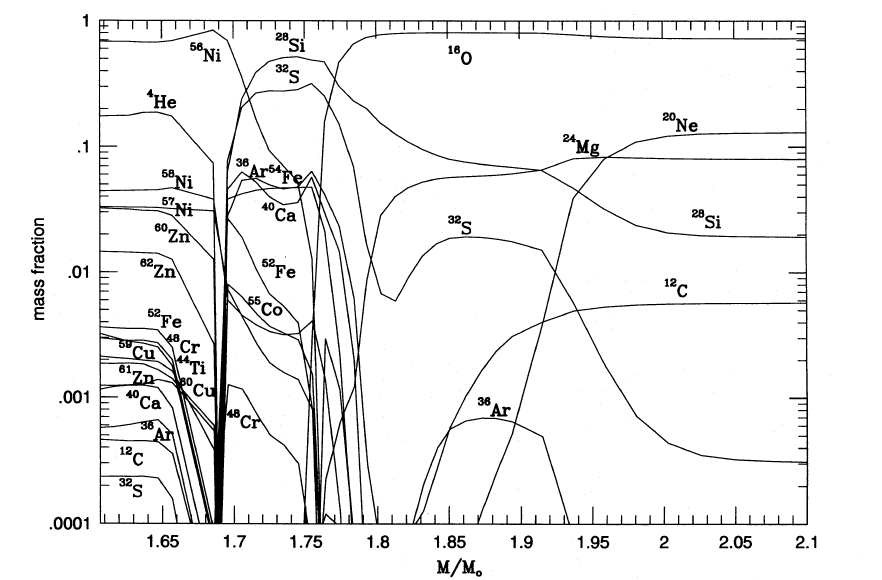}
    \caption{Nucleosynthesis products (including unstable nuclei before decay) of a core-collapse supernova as a function of the radial mass coordinate for a 20\,M$_\odot$ star from \cite{Thielemann.Nomoto.Hashimoto:1996}.}
    \label{fig:ccex20}
\end{figure}

The Pearson correlation between Fe and Mg through Ti is robust and significantly different from zero (PCC$>$0.7-0.8). This reflects an origin of these elements alongside that of Fe, which predominantly originate from massive stars and their core-collapse supernovae. Massive stars that undergo carbon burning will pass through all hydrostatic burning stages until core-collapse. This eventually leads to core-collapse supernova explosions during which the Fe-group elements are produced in explosive Si-burning in the innermost ejected mass zones, along with products of explosive C, Ne, and O burning during the passage of the supernova shock wave through these layers (see Fig.~\ref{fig:ccex20}). 

This also includes the somewhat unaltered ejection of the outer layers of the star that contain products from hydrostatic He and C burning. Overall, this combines the ejection of Fe and Mg (with a strong correlation), but also O, which is typically considered as an $\alpha$-element as well. We will later discuss the peculiar behavior of O, together with that of C and N, as this leads to a division into regular and CEMP stars.


\begin{figure}[h!]
\centering
\includegraphics[width=0.85\linewidth]{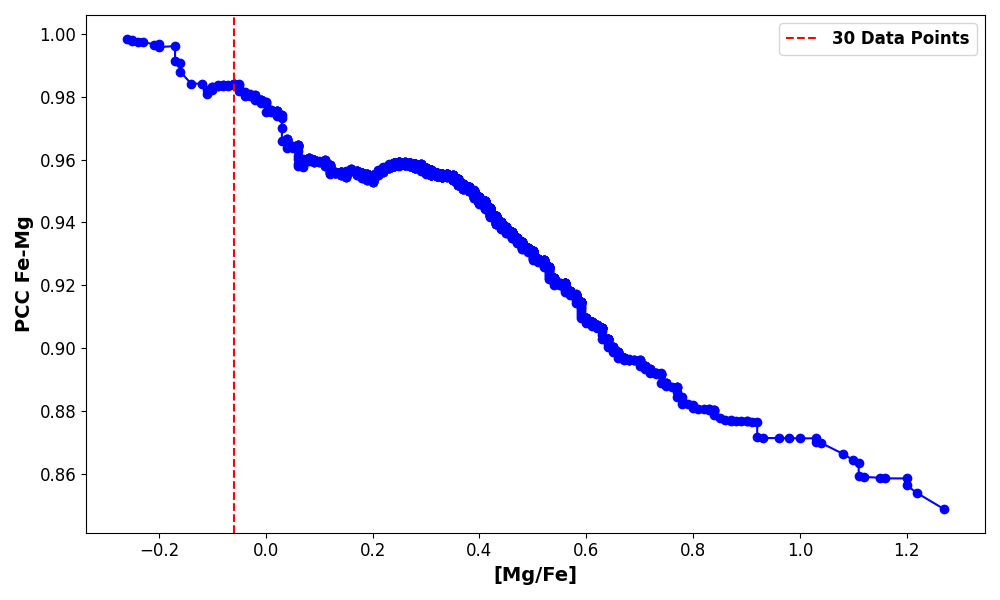}
\caption{Exploration of the integrated Pearson correlation between Fe and Mg for varying levels of [Mg/Fe] enhancement. The plot shows for all stars the  PCC values integrated up to a given [Mg/Fe] ratio. Values range up to or even beyond 1 for extremely metal-poor stars (see e.g. \citealt{Francois.ea:2020} for stars with [Fe/H] $\sim-$4). 
Depending on the stellar progenitor mass, hydrostatic C-burning (which is responsible for most of the Mg ejected in core-collapse supernovae) will result in varying amounts of Mg. This leads to a variation in [Mg/Fe]. This is as shown for the overall sample of stars utilized for this analysis (see Table \ref{tab:correlation_resultsB}) which includes all low-metallicity stars with observed Mg and Fe abundances. The full range of  [Mg/Fe] ratios, including also the high [Mg/Fe] tail, leads to a high correlation coefficient above 0.8. This strongly indicates the co-production of Mg with Fe and thus a core-collapse supernova origin. The dashed red line marks the threshold of 30 data points beyond which such a statistical analysis is meaningful.}
\label{fig:FeMg.png}
\end{figure}

The statistical relations found for the $\alpha$-elements with Fe are consistent with expectations from theoretical stellar and supernova modeling for the production of $\alpha$-elements (Mg to Ti) in core-collapse supernovae of massive stars  \citep{matteucci86,Timmes.Woosley.Weaver:1995}). 

As observed in low-metallicity stars, $\alpha$-elements display enhanced [$\alpha$/Fe] ratio (i.e., supersolar values at a typical level of 0.4), because $\alpha$- elements are mainly produced in progenitor core-collapse supernovae prior to the stars' births. 
At higher metallicities of [Fe/H] $\gtrsim-1.5$, stars begin to show lower [$\alpha$/Fe] ratio, culminating at 0.0 for stars with solar metallicities. This change can be attributed to increasingly dominant contributions of Fe by type Ia supernovae as part of the later stages of galactic evolution, while CCSNe continue to contribute only between 1/3 and 1/2 to e.g. the solar abundance of Fe. This behavior has been confirmed with ample chemical evolution models and stellar explosion calculations \citep[see e.g.][]{Nomoto.Tominaga.ea:2006,Woosley.Heger:2007,Sukhbold.Ertl.ea:2016,Thielemann.ea:2018,limongi18,Curtis.ea:2019,Ebinger.Curtis.ea:2020}. While this behavior is of great interest with regard to understanding general galactic evolution, for the metallicity range considered here, [Fe/H]$<-$2, the contributions by type Ia supernovae are taken as irrelevant.

In Figure~\ref{fig:ccex20}, the abundances resulting in the explosion of a 20\,M$_\odot$ star are shown. The main products of explosive burning are produced essentially from a fuel composition of symmetric nuclei with $N\approx Z$. While $^{24}$Mg stems dominantly from outer zones, which experienced prior hydrostatic C-burning during stellar evolution, and explosive Ne-burning contributing minor amounts of the $\alpha$ nuclei $^{28}$Si, $^{32}$S, and $^{36}$Ar, the main contribution to $^{28}$Si, $^{32}$S, $^{36}$Ar, and $^{40}$Ca comes from explosive O-burning. Incomplete and complete explosive Si-burning contribute $^{44}$Ti, $^{48}$Cr, $^{52}$Fe, $^{56}$Ni, and $^{60}$Zn, which are all $N=Z$ (but unstable) nuclei, decaying to $^{44}$Ca, $^{48}$Ti, $^{52}$Cr, $^{56}$Fe, and $^{60}$Ni. This clearly relates the production of the classical $\alpha$ nuclei, that make up elements Si, S, Ar, Ca, and Ti, to Fe (with Fe stemming from core-collapse supernovae during the early phase of galactic evolution). We will discuss the correlation of the Fe-group nuclei beyond Cr in the next Section.


Considering Mg as an example of elements which stem from hydrostatic burning zones, large variations in the mass of the hydrostatic zones as a function of stellar mass can lead to large variations in ejecta, while explosive burning ejecta vary by a smaller amount. 
Typical observed [Mg/Fe] ratios are of the order 0.4-0.6 \citep[see e.g.][]{Francois.ea:2020}. Fig.~\ref{fig:FeMg.png} shows that for the sample of all stars with [Mg/Fe] $<$ 0.6, the PCC is as high as 0.9. When including all observations, which can reach up to [Mg/Fe] $\sim$ 1, the overall PCC still amounts to 0.8, as given in Table ~\ref{tab:correlation_resultsB}. This strong correlation, despite relatively large variations found in Mg ejecta, goes back to the fact that the more massive core-collapse supernova progenitors, with large hydrostatic burning zones, also experience stronger explosions with larger $^{56}$Ni ejecta \citep[as found in self-consistent 3D supernova models, see e.g.][]{Burrows.Radice.ea:2020,Vartanyan.ea:2022,Burrows.ea:2024,Wang.BurrowsG:2024}.

\section{The correlation of Fe with Fe-group nuclei}
\begin{figure}[h!]
    \centering
    \includegraphics[width=0.53\linewidth]{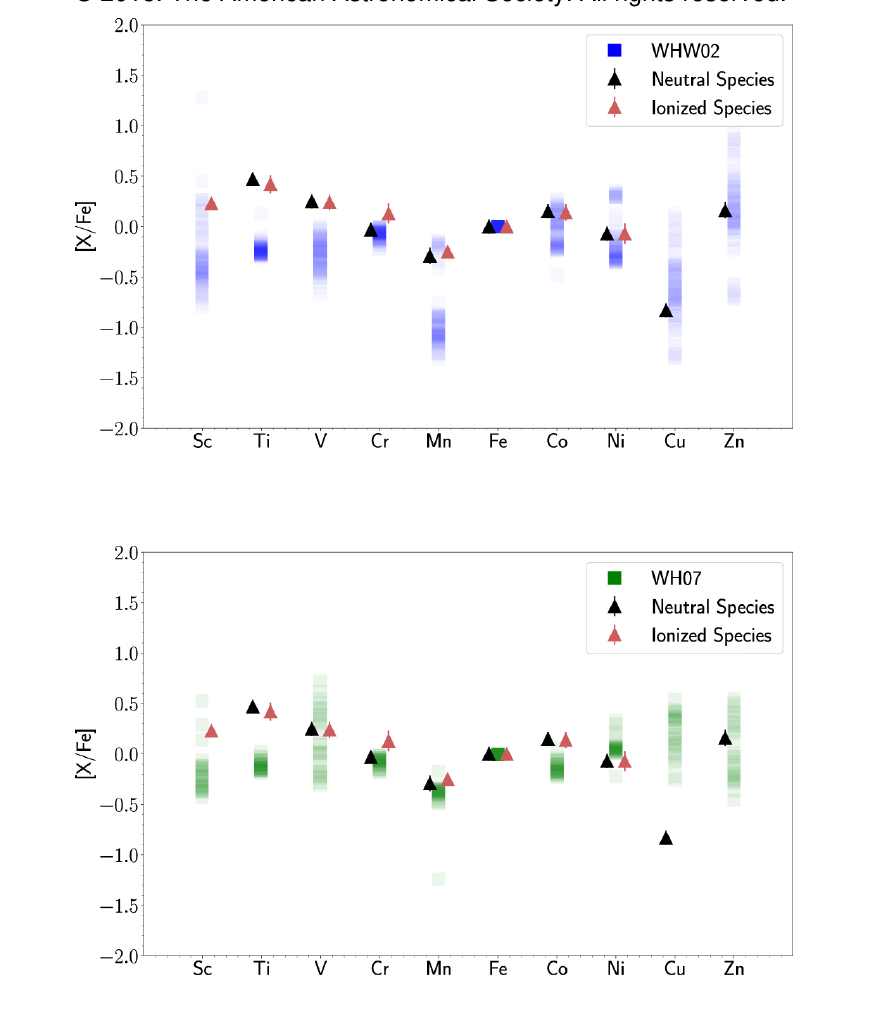}
    \caption{Top and bottom panels: Observed abundances for the low-metallicity ([Fe/H] = $-$2.3) star HD 84937 \citep{Sneden.ea:2016} with extended coverage of Fe-group elements (triangles) compared to predicted Fe-group results from core-collapse supernova studies by \cite{Curtis.ea:2019} for two sets of massive progenitor stars, WHW02 \citep{Woosley.Heger.Weaver:2002} and WH07 \citep{Woosley.Heger:2007}. The transparency of the bars is normalized such that a darker color indicates more stellar models producing a particular element ratio. Integrated results over all models are given in the solid color as shown in the legend.}
    \label{fig:GhoshFe}
\end{figure}

The correlations of Fe with all elements within the Fe-group from Cr to Zn are robust and significantly different from 0, as can be seen in Table \ref{tab:correlation_resultsB}.
The calculations for Fig.~\ref{fig:ccex20} did not consider the weak interactions in the inner ejected zones during collapse and explosion, including electron captures at high densities along with neutrino absorption during the explosion. The latter leads to a slightly proton-rich environment with $Y_e>0.5$, as can be seen in Fig.~10 of \cite{Ghosh.ea:2022}. 

\begin{figure}[h!]
    \centering
    \includegraphics[width=0.6\linewidth]{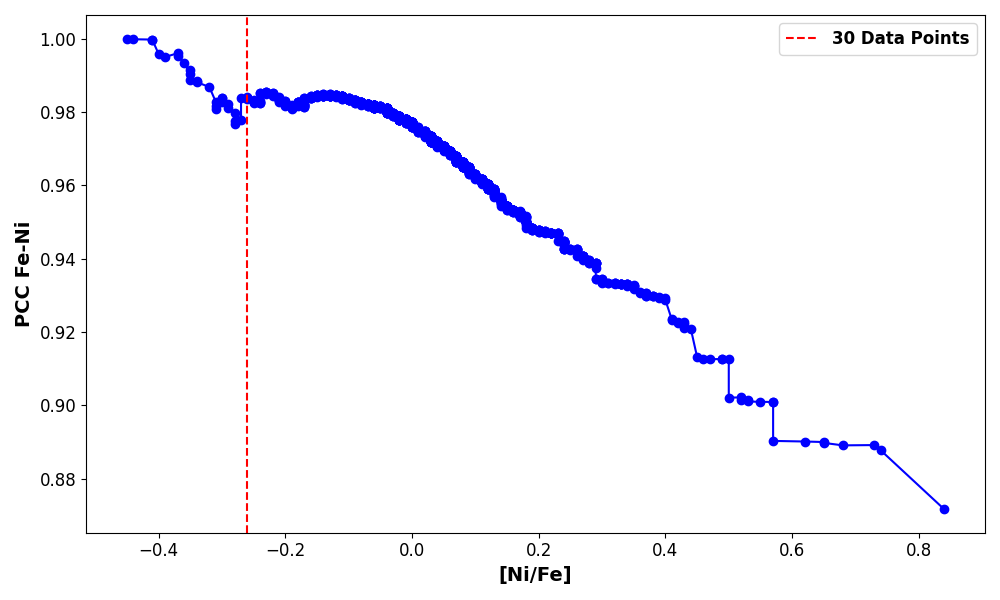}
    \caption{Exploring the Pearson correlation between Fe and Ni across varying levels of observed [Ni/Fe] abundances. Contrary to Mg (see Fig.~\ref{fig:FeMg.png}), Ni is solely produced in explosive Si-burning (jointly with Fe). Hence, the theoretically predicted variations among the progenitor masses of core-collapse supernovae are smaller. This is consistent with observations, which show a very high correlation between Ni and Fe among the whole range of [Ni/Fe], as indicated by the large PCC values. This is supported by Fig.~\ref{fig:GhoshFe}, which shows only small variations within the entire Fe-group for the entire grid of stellar progenitor masses and metallicities.}
    \label{fig:FeNi}
\end{figure}

Correctly considering  the role of weak interactions during the collapse and explosion has considerably improved the agreement with observations of Fe-group elements in low-metallicity stars \citep[see e.g.][]{Curtis.ea:2019,Ghosh.ea:2022}, which suffered from discrepancies before. This permits that also nuclei 
such as $^{64}$Ge (decaying to $^{64}$Zn), are produced
when considering the weak interaction effects in core-collapse supernovae. Therefore, all abundance patterns of Fe-group elements up to Zn observed in low-metallicity stars \citep{Cowan.ea:2020,Sneden.ea:2023}
can be well reproduced by CCSNe contributions alone \citep{Curtis.ea:2019,Ghosh.ea:2022}, as can be seen in Fig.~\ref{fig:GhoshFe}. This Figure spans the whole range of nuclei produced in explosive Si-burning from Sc to Zn and clearly shows that a good weak interaction treatment can reproduce the abundance pattern of the entire Fe group.

Fe and Ni both stem from explosive Si-burning (contrary to Fe vs. Mg with a contribution from hydrostatic burning zones, varying strongly with progenitor mass, as well as explosive burning). Therefore, we find a smaller variation in the Fe/Ni ratio among the entire sample of observed low-metallicity stars, which points directly to the abundance pattern of ejected matter from contributing progenitor stars. This is shown in Fig.~\ref{fig:FeNi}, and leads also to a high correlation between these elements.

\section{Suggested sources for contributions to elements beyond Fe}
As a preparation for discussions related to elements heavier than Fe, we want to list here in advance the processes that could be responsible for their production. This will facilitate to address them when continuing the correlation analysis across the Periodic Table of the elements. In addition to the major processes, the s- and the r-process of slow and rapid neutron capture, observers also introduced a still not understood LEPP process, which stands for Light (heavy) Elements Primary Process \citep{Travaglio.Gallino.ea:2004}. It was mostly attributed to abundances of Sr, Y, and Zr elements, usually made in the s-process. However, because of its dominant origin from low- and intermediate-mass stars, it is only seen in chemical evolution at metallicities above [Fe/H]=-2 to -1.5. Whether the core-collapse supernova related $\nu$p-process, which can produce nuclei up to $A\approx80-90$ \citep{Froehlich.Martinez-Pinedo.ea:2006}, or an early limited r-process can explain this abundance feature is still debated. 

\begin{figure}[h!]
    \centering
    \includegraphics[width=0.95\linewidth]{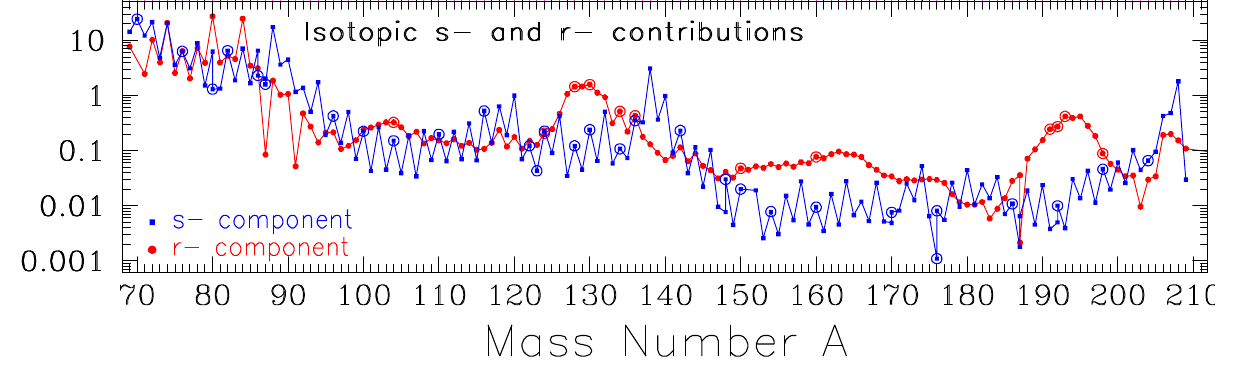}
    \caption{Decomposition of solar-system heavy element abundances in their s- and r-process fractions; abundances are scaled arbitrarily, from \cite{Prantzos.ea:2020}. Notice that what is called here the r-process component is in fact the difference between solar abundances and the s-process component. In the range up to $A\approx$90 this might include contributions by the $\nu$p-process and charged-particle freeze-out in r-process conditions, which is underlined by a deviation from the otherwise smooth r-process pattern.}
    \label{fig:Prantzossr}
\end{figure}

The division of solar s- and r-process abundances is based on first obtaining the s-process composition from fitting abundances of s-only nuclei (those nuclei in the nuclear chart which can only be made by the s-process, because they are blocked from r-process contributions via stable more neutron-rich isobars). The s-process abundances are explained by a superposition, either of components with different neutron exposures or by integrating stellar ejecta contributions over stellar initial masses and their metallicities, because the s-process acts via neutron captures on pre-existing Fe \citep[see e.g.][]{Kaeppeler.Gallino.ea:2011}. In a second step, this s-process composition is subtracted from solar abundances. The most recent version of this decomposition by \cite{Prantzos.ea:2020} is shown in Fig.~\ref{fig:Prantzossr}.

It is also expected that the solar r-process abundances are a superposition of ejecta components. The observed r-process abundances, which show dominant contributions at the metallicities of interest in this analysis, have been divided by observers into subclasses of observed abundance patterns: limited-r, r-I, and r-II, which show different enhancements of Eu in comparison to Fe. With stable isotopes $^{151}$Eu and $^{153}$Eu, Eu is dominated by the r-process (see Fig.~\ref{fig:Prantzossr}). The different subcategories of Eu enhancement, introduced above, are determined by their [Eu/Fe] ratios: 
limited-r stars with [Sr/Ba]$>$0.5 and [Eu/Fe]$<$0.3, 
r-I stars with 0.3$<$[Eu/Fe]$<$0.7, and 
r-II stars with [Eu/Fe]$>$0.7 \citep{Holmbeck.Hansen.ea:2020}. 
We will discuss in later Sections whether these subcategories are related to different r-process sites/contributions, which have been introduced as limited-r, main-r, and robust-r by \cite{Frebel:2018}, or as weak-r and strong-r with or without an actinide boost contribution, i.e. unusually large Th and/or U abundances compared to the enhancements of other heavy elements such as Eu \citep[e.g.][]{Cowan.Sneden.ea:2021}. It has turned out for quite a while that the observed r-I and r-II abundances  have a universal pattern between the lanthanides ($A>140$) and the third r-process peak at $A=195$ \citep[a typical feature of the main r-process,][]{Sneden.ea:2000}, while in a number of r-II stars also an actinide boost, has been observed (in addition to a robust/main r-process abundance pattern). \cite{Roederer.ea:2022} have noticed such a universal pattern also between between Se ($A=80$) or Sr ($A=88$) and the second r-process peak at Te ($A=130$), however, scaled in comparison to the main r-process abundances, with the scaling depending on the  categories defined above. 

Having presented these preparatory comments, we will continue with our correlation analysis across the Periodic Table in the following sections, with the main emphasis to find out if such abundance patterns of heavy elements are correlated with Fe, hinting at a strong, moderate, or no co-production of these elements with Fe in core-collapse supernova events.

\section{The correlation of Fe with elements up to Se}

\begin{figure}[h!]
    \centering
    \includegraphics[width=0.8\linewidth]{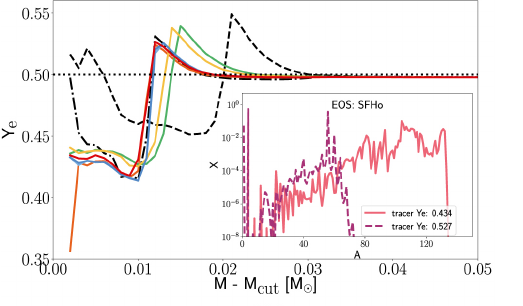}
    \caption{Electron faction-distributions, $Y_e$, in the innermost ejected zones for different high-density equations of state, indicated by differently color-coded lines inside $(M-M_{cut})/M_\odot$=0.03 \citep[see details in][]{Ghosh.ea:2022}. These 1D spherical explosion simulations utilized an improved treatment of weak interactions during collapse and explosion. The very innermost zones contain matter with $Y_e$$<$0.5, due to electron-captures during the collapse. The adjacent zones result in $Y_e$$>$0.5, based on neutrino and anti-neutrino captures on neutrons and protons. The outer zones show $Y_e$$\approx$0.5, as a result of earlier hydrostatic evolution phases. The insert displays the results of a $\nu$p-process in the zones with $Y_e$$>$0.5 as well as a very weak r-process in the zones with $Y_e$$<$0.5.}
    \label{fig:nup}
\end{figure}

In the previous sections we have found high correlations for the $\alpha$-elements from Mg to Ti with Fe, as well as for all Fe-group elements from Cr to Zn, which are all shown to be of core-collapse supernova origin. These high correlations continue up to Se or $A\approx 80$. \cite{Froehlich.Martinez-Pinedo.ea:2006} discovered that in the core-collapse environment, with a strong neutrino and anti-neutrino flux from the collapsed core, the $\nu$p-process can operate which permits even overcoming the long beta-decay of $^{64}$Ge, where proton-captures in slightly proton-rich conditions stop. Fig.~\ref{fig:nup} from \cite{Ghosh.ea:2022} shows the electron fraction, $Y_e$, distribution as a function of radial mass outside of the mass cut (matter inside the mass cut will fall onto the nascent neutron star and not get ejected). While the results depend somewhat on the utilized nuclear equation of state of high density matter (which controls the bounce after core-collapse), regions with $Y_e>0.5$ are always found due to the neutrino interactions, leading to the $\nu$p-process and the production of elements up to Se. 

This leads to an environment where the antineutrino capture on free protons causes a non-negligible production of neutrons, enabling an $(n,p)$ reaction on $^{64}$Ge, which serves like a beta decay and permits further proton captures and fast beta decays up to the $A\approx 80$ region. This can also be seen in the insert of Fig.~\ref{fig:nup} from \cite{Ghosh.ea:2022} for $Y_e=0.527$ which presents the results of this $\nu$p-process. It produces elements up to Se in regular core-collapse supernovae. This is consistent with the high correlations of Fe with Ge and Se as seen in Table \ref{tab:correlation_resultsB}. However, we note that for Ge and Se the number of stars with measured abundances is rather small; hence, despite the relatively small p-values, these results have to be considered somewhat uncertain. Additional data would help confirm these findings.
Moreover, it is still an open question whether the slightly neutron-rich very innermost zones will be ejected in self-consistent 3D simulations. If so, this could result in a very limited r-process, producing only elements up to the second r-process peak (but not beyond).

\section{Speculation on Sr through Te}

In Table~\ref{tab:correlation_resultsB}, the abundances from Sr to Ag show typical PCC values of the order 0.1 to 0.2, representing a weak correlation. This would argue for a partial co-production with Fe, but dominant contributions have to come from other (non-CCSN) sources. However, rather surprisingly high PCC-values are found for Rh, Pd, and Ag. We caution that these are based on small datasets which call into question the significance of any of these moderately high correlations. When checking Fig.~\ref{fig:nup}, discussed in the previous Section, a very limited r-process is indicated with nucleosynthesis occurring up to the second r-process peak around $A$ = 130. When considering the r/s-decomposition of heavy elements for the presolar nebula material, shown in Fig.~\ref{fig:Prantzossr}, one sees in addition to the prominent s- and r-process peaks, regions where the r-process dominates, e.g. just for the mass range directly above $A$=100 (including the elements Rh, Pd, and Ag), as well as around $A$=150-170, which includes the element Eu. If elements from a very weak/limited r-process, as shown in Fig.~\ref{fig:nup}, would be ejected in regular CCSNe, they would be correlated with Fe via a strong or moderately strong correlation. This seemed to be first supported by the observational data for Pd (and possibly also Rh and Ag) but the position of thelower metallicity cutoff at [Fe/H]$=-2$ might question this finding.


A production of nuclei beyond the $\nu$p-process region and up to the second r-process peak has previously been suggested, as is indicated in the insert of Fig.~\ref{fig:nup} taken from \citep{Ghosh.ea:2022}. It shows an abundance pattern commensurate with the innermost ejected layers with a low $Y_e$, due to electron-captures in dense matter during core-collapse. While there exists a chance that these zones survived the enhancement of $Y_e$ by neutrinos during the explosion, such effects need to be further tested with self-consistent 3D simulations. If confirmed, one would expect a correlation of these elements with Fe, similar to those shown for  Rh, Pd, and Ag, but this should apply to the entire mass region from $A$=80 to 130, which is not the case. Unfortunately, observational data of elements in the second r-process peak (Xe, Te) are very scarce, because they are based on UV high-resolution spectroscopy. Thus, a correlation analysis is hampered by very limited datasets. 

Given this somewhat uncertain situation, we decided to move the stellar metallicity cut-off utilized in our heavy element analysis from [Fe/H] = $-2$ to $-2.5$. Reducing the cutoff helps with eliminating stars with s-process element contributions (they exist to a minor extent already for $-$2.5$<$[Fe/H]$<-$2) from the samples used. When applying this procedure, it should reduce possible s-process contributions to the PCC values for elements in the region $A$=80 to 100, and would leave the higher correlations for $A$ just beyond 100, where an r-process sub-peak dominates (see Fig.\ref{fig:Prantzossr}) intact. However, after making this cut, the PCC values for Sr through Mo were even slightly reduced, and the moderately high PCC values for Rh, Pd, and Ag disappeared, showing once again that a small number of data is problematic, and that  the statistical significance of these results, at face value, needs to be questioned.

Thus, the result of this exercise is that we expect for the elements from Sr through Ag small PCC values of the order of 0.1 to at most 0.2.


\section{The correlation of Fe with nuclei from Sr to Eu and possibly Yb}

\begin{figure}[h]
    \centering
    \includegraphics[width=0.8\linewidth]{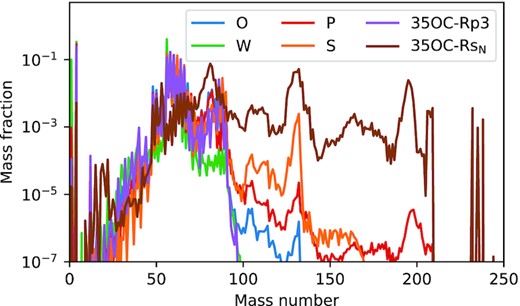}
    \caption{Stellar rotation and the strength of magnetic fields determine the strength of an r-process in magneto-rotational supernovae. Matter with small $Y_e$, neutronized via electron-captures during the collapse, can be ejected (due to the fast winding of magnetic fields) before neutrino and anti-neutrino-captures would otherwise again lead to an increase of $Y_e$. This Figure from \cite{Reichert.ea:2023} shows  that nuclei beyond the second r-process peak at A=130 can be produced. But when a realistic parameter space of rotation and magnetic fields is assumed only a weak r-process is realized that nevertheless produces elements such as Eu and beyond.}
    \label{fig:Reichertweak}
\end{figure}

After the conclusion of the previous Section about finding weak correlations from Sr through Ag, the overall result for the entire region from Sr to Eu, or possibly Yb, is that only weak correlations in the range of 0.1 to 0.2 are derived. This points to a contribution or component of these elements that is co-produced with Fe. But a significant fraction of these elements must also stem from other sources that are not related to core-collapse supernovae. In the literature, a number of predictions for a rare class of magneto-rotational core-collapse supernovae are made, which depend on rotation and magnetic fields. These can experience a weak or limited r-process in their ejecta which seems to go beyond the second r-process peak and can cause a moderate production of elements such as Eu, even extending up to Yb, Lu, and Hf \citep{Winteler.Kaeppeli.ea:2012,Moesta.ea:2015,Nishimura.Takiwaki.Thielemann:2015,Nishimura.Sawai.ea:2017,Reichert.Obergaulinger.ea:2021,Reichert.ea:2023}. Thus, a fraction of core-collapse supernovae can produce the elements from Sr to Eu and possibly Yb, Lu, and Hf. This leads to a correlation of these elements with the production of Fe.
For the subset of limited-r stars, this correlation is actually strong, as supported by the high PCC value of Eu for these stars. But this r-process contribution from magneto-rotational supernovae that reaches beyond the second r-process peak (producing Fe and also rare earth elements), needs to be complemented by a strong r-process source. This would explain the overall smaller PCC values for the entire element range from Sr to Eu, and possibly Y, Lu, and Hf observed in low-metallicity stars. If a main source would come with a lack of a correlation with Fe, this would lead overall to relatively small (but still significant) correlations with Fe for this element range \citep[see also][]{Farouqi.Thielemann.ea:2022}.

\section{The correlation of Fe with heavy nuclei up to the actinides Th and U}

Here we come to a nuclear mass region which can only be produced in a main r-process without significant co-production of Fe as it occurs in CCSNe. Two major astrophysical sites with suitable environments exist: 

\begin{figure}[h!]
    \centering
    \includegraphics[width=\linewidth]{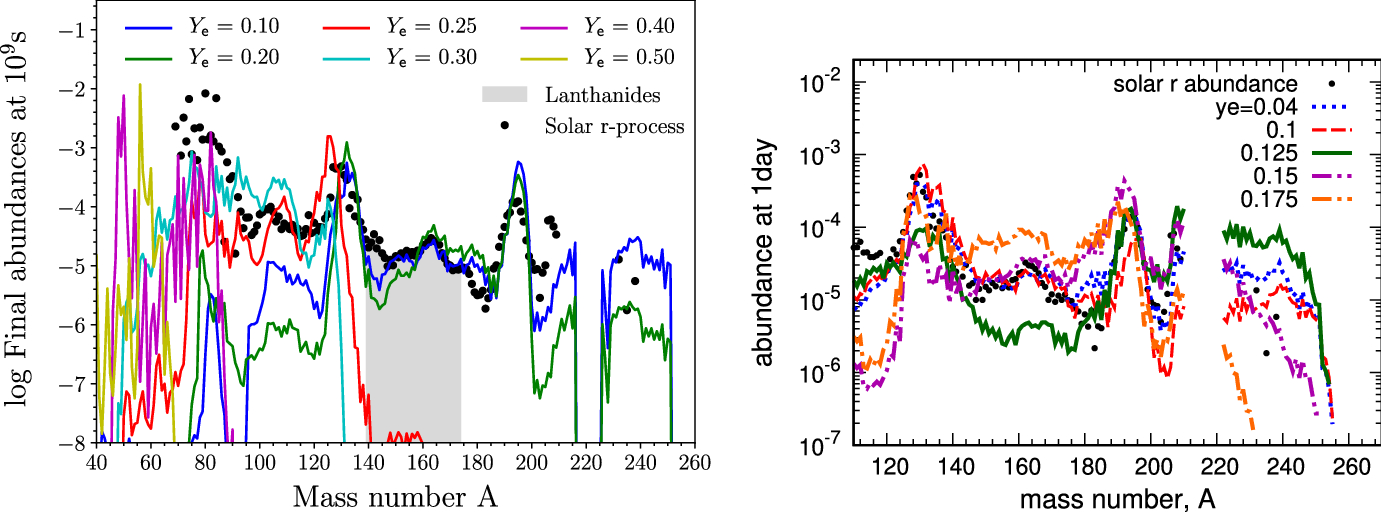}
    \caption{(Left panel) Isotopic abundances as a function of the mass number A at $10^9$
 seconds after a compact binary merger for trajectories characterized by $Y_e$ \citep{Perego.Thielemann.Cescutti:2021}. Black dots represent the Solar r-process residuals. (Right panel) Isotopic abundances as a function of A \citep{Thielemann.ea:2020}, focusing on low 
 $Y_e$ values and the abundances of actinides.}
    \label{fig:ArThistrongr}
\end{figure}

(1) Collapsars/Hypernovae:\citep{fujimoto07,Ono.Hashimoto.ea:2012,Siegel:2019,Siegel:2022,Just.Aloy.ea:2022,Dean.Fernandez:2024,Issa.Gottlieb.ea:2024} originate from massive stars with a core that finally collapses into a black hole, yet (for fast rotation and emerging magnetic fields) nuclei are produced in the black-hole accretion disk outflows without significant amounts of Fe. As the ratio [r/Fe] is orders of magnitude higher than the solar values, a correlation with Fe seems negligible. While there exists an extended literature suggesting collapsars/hypernovae as a viable r-process site, a final observational proof is still missing \citep{Anand.ea:2024}.

(2) An alternative explanation is provided by compact binary mergers
\citep[see e.g.][]{Thielemann.Eichler.ea:2017,Bauswein.Just.ea:2017, rosswog18,Cowan.Sneden.ea:2021,Perego.Thielemann.Cescutti:2021,Wanajo.Hirai.Prantzos:2021,Just.Vijayan.ea:2023,Shibata:2023,}. While early inhomogeneous galactic evolution simulations only predict any potential impact above metallicities of [Fe/H] $\sim-$2.5 \citep{wehmeyer15,Wehmeyer.ea:2019}, further developments of such codes could potentially identify their existence and impact down to [Fe/H] $\sim-$3 \citep{VandeVoort.ea:2022}.  This has also been indicated as a possibility by \cite{Farouqi.Thielemann.ea:2022} who showed the appearance of r-process r-I and r-II stars already at such low metallicities (see also Fig.~\ref{fig:EuHEuFe}). 

Fig.~\ref{fig:ArThistrongr} shows two r-process simulations as a function of $Y_e$ in ejected matter, which relate to both scenarios (1) and (2). The right panel focuses on the resulting actinide abundances, permitting a so-called actinide boost.
How actinide boosts (often found in the r-process abundance patterns of low-metallicity r-II stars) are explained, whether requiring different origins than r-I abundance patterns, or whether the boost seen in r-II patterns relates either to compact mergers vs. collapsars, or binary neutron star mergers vs. neutron star black hole mergers, remains an open question \citep{Wanajo.Fujibayashi.ea:2024}.

In Section \ref{onset}, we will discuss the global correlation behavior of all low-metallicity stars combined with the overall time and metallicity evolution. This could point to the onset of different strong r-process sources.

\section{Correlation of Fe with elements from C to Na}

\begin{figure}[h!]
\centerline{
\includegraphics[width=0.9\linewidth]{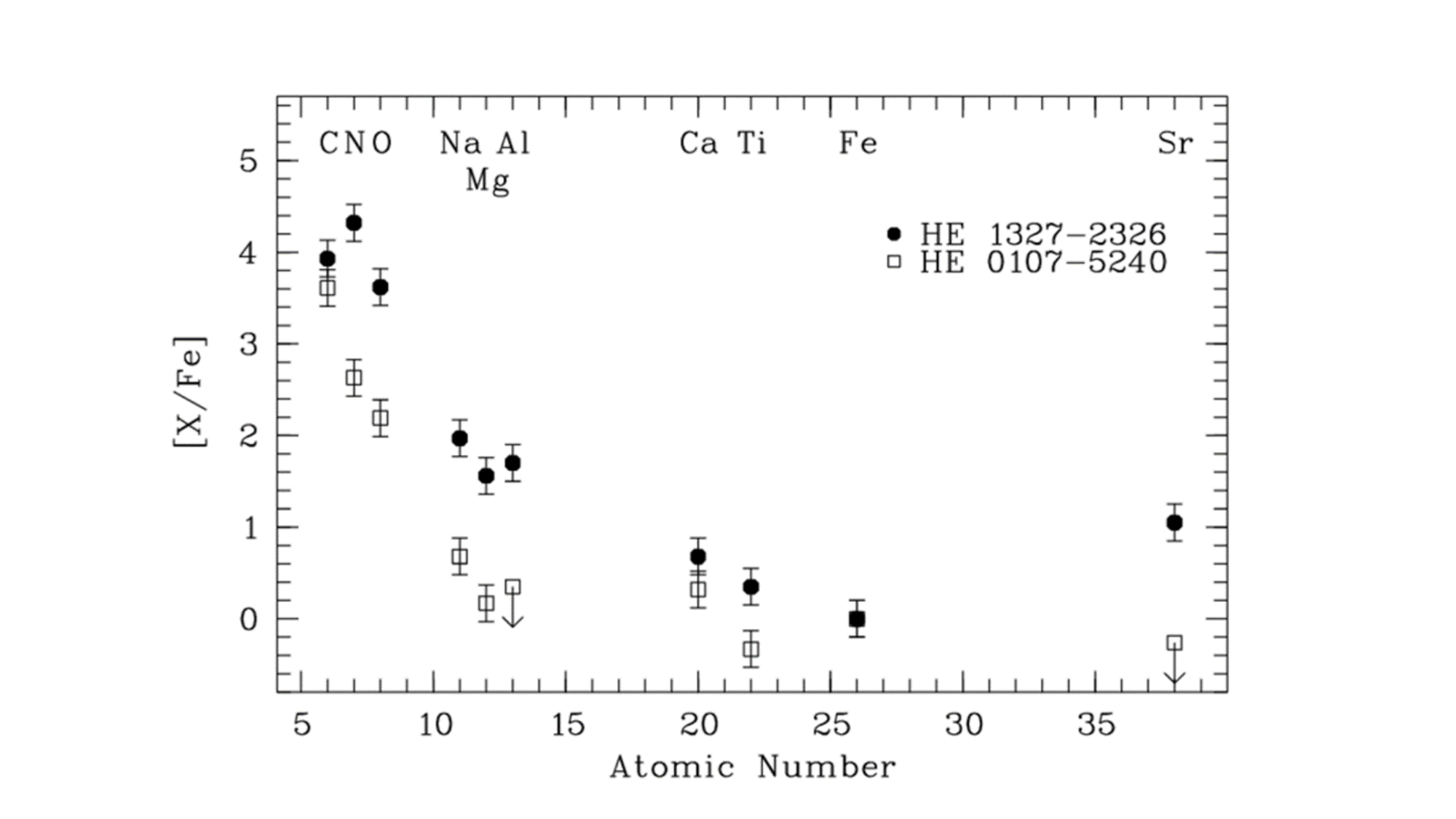}}
\caption{Two examples of very carbon enhanced metal-poor (CEMP) stars from \cite{Frebel.ea:2005} with [C/Fe]$\approx$4, showing also high N and O enhancements beyond 2, while heavier elements such as Fe are strongly suppressed in comparison to CNO nuclei.}
\label{figCemp}
\end{figure}

We will discuss here the fact that the correlation with Fe is very weak for C, N, and O up to Na, when considering the whole set of EMP stars. There exist subsets, characterized by strong C, N, or O enhancement in comparison to Fe (Fig.~\ref{figCemp} shows two stars belonging to this subset
introduced by \cite{Frebel.ea:2005}, see also \cite{Frebel:2024}. They belong to the class of CEMP (carbon enhanced metal-poor) stars which show C (and/or) N, O, and Na enhancements over intermediate-mass and Fe-group elements.

 In a series of papers \citep[see e.g.][]{Nomoto.Tominaga.ea:2006,Nomoto.ea:2013,Leung.Nomoto:2024}, the group around Ken Nomoto has advocated for explosions of massive stars, which will eventually form black holes \citep[see also][]{Ji.ea:2024}. The hydrodynamic instabilities in the central zones can (during the explosion mechanism) lead to an extended mixing of material from the innermost zones to outer (later ejected) layers, while most of the inner layers end in the newly formed black hole. Such explosions, with low $^{56}$Ni ejecta masses, are very faint and of low energy. Hence, they have been termed faint SNe. However, the extended mixing provides some (small) contribution of Fe-group elements to the outer ejected zones that consist mainly of matter from the earlier hydrostatic burning stages of H, He, and potentially C-burning. 
 Nucleosynthesis in faint SNe is characterized by a large amount of fall-back matter, which can explain the abundance patterns found in the most Fe-poor stars that are characterized by very low Fe and comparably large C abundances. 

 Here we analyze the observed correlation coefficients between Fe and C abundances in metal-poor stars, particularly when considering the distinction between the Carbon-Enhanced Metal-Poor (CEMP) and the remaining stars: (1) We find a weak, but overall significantly different from zero correlation of 0.1 when taking the whole set of EMP stars. This indicates that on average, there is a slight, and statistically significant, tendency for the abundances of Fe and C to vary together in stars. However, the strength of this relationship is relatively weak. (2) A somewhat stronger (but still weak) correlation of the order of 0.25 is found for the subset of CEMP stars alone (those with [C/Fe] $>$ 0.7). 
 This suggests a somewhat stronger association between Fe and C abundances within this specific population of stars (related to the progenitor faint supernovae, as discussed above), which is, however, affected by two competing effects. (i) For a given type of progenitor faint supernova the C/Fe ratio is fixed, only the dilution caused by mixing its ejecta into a nearby, or not-so nearby, star-forming gas cloud leads to a variation of the total amount of contributed C and Fe. Nevertheless, the C/Fe is identical for different star forming clouds, which would lead to a strong linear correlation. (ii) However, the observed fact is that the number of CEMP stars increases strongly with decreasing metallicity, i.e. Fe abundance. This effect of an increasing number of CEMP stars with a decreasing Fe abundance dampens the otherwise expected strong correlation. (3) A strong correlation of the order of 0.7 is found within the subset of stars with [C/Fe] $<$ 0.7. This indicates a much stronger relationship between Fe and C abundances among those stars in comparison to the CEMP stars. Regarding the progenitors of this population, which we label "regular supernovae", variations in Fe yields are strongly correlated with the corresponding variations in C ejecta.
 
These findings suggest that while there is a weak but overall significant  correlation between Fe and C abundances across all stars, the strength of this correlation varies significantly, depending on their level of carbon enhancement. We interpret the stronger correlations observed for the subset of stars with [C/Fe] $<$ 0.7 which make up the majority of VMP stars, as the result of regular core-collapse supernovae.  On the other hand, the CEMP stars seem to be found mostly at the lowest metallicities ([Fe/H] $<-$3.0) and likely reflect the nucleosynthesis products of the first generation of massive black-hole forming zero-metallicity stars in the Universe. 

\begin{figure}[h!]
\centering
\includegraphics[width=0.48\linewidth]{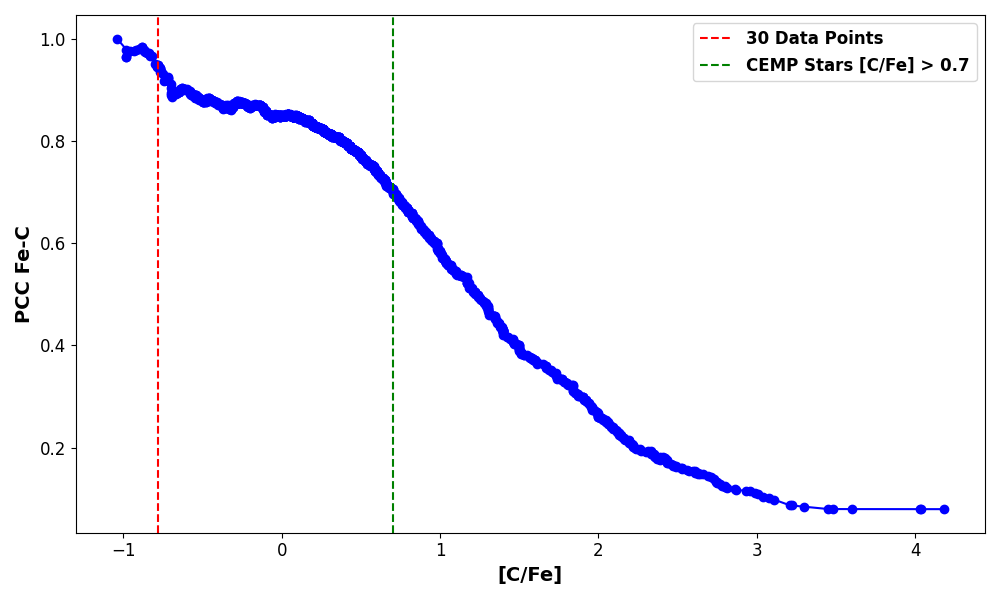}
\includegraphics[width=0.48\linewidth]{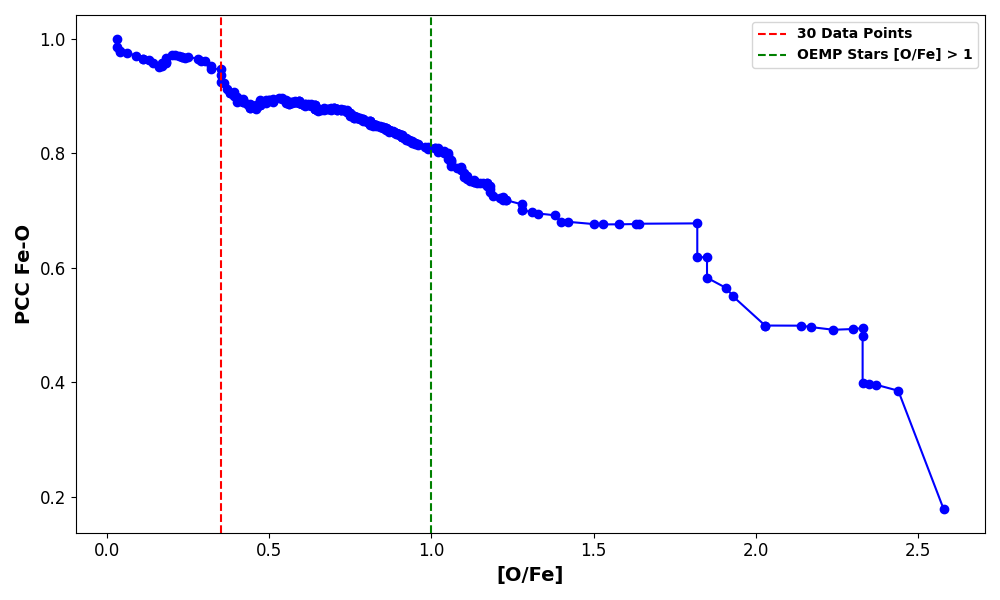}
\caption{Exploring the Pearson correlation between Fe and C as well as O across varying levels of [C/Fe] and [O/Fe] enhancements. For "regular" stars with [C/Fe] or [O/Fe]$<$1 the correlations with Fe are high and point to a supernova co-production of C and O along with Fe in the progenitors. On the other hand, for CEMP stars with large C/Fe and/or O/Fe enhancements, the correlation with Fe decreases strongly, pointing to contributions from other exploding objects (faint/failed supernovae where Fe is accreted onto the nascent black hole) with vanishing amounts of Fe ejecta.
The dashed red line depicts the critical limit of 30 stars in the statistical data set. We also include a green dashed line which marks the limit between C or O enhanced CEMP and OEMP stars and regular stars, pointing to existence of special faint/fallback supernova sites with small to negligible Fe ejecta.}
\label{fig:FeCNa}
\end{figure}

We also find similar tendencies for the correlation between Fe and N, which is weak (around 0.2) but significantly different from zero. However, when separating the data into NEMP stars (with [N/Fe] $>$ 1) and the non-NEMP stars, the correlation between Fe and N within the NEMP class is 0.36, while within the non-NEMP class, it is 0.72. This finding parallels what we find with regard to C. Both elements (C and N) appear to exhibit a similar behavior.
The same also holds true for the correlation between Fe and O, which is weak but still significantly different from zero, with a coefficient of only 0.16. However, when the data are divided into two groups: oxygen-rich metal-poor stars ([O/Fe] $>$ 1) and regular stars ([O/Fe] $<$ 1), the correlation between Fe and O within the first group is 0.21, while it is 0.80 for the second group. These results mirror our earlier findings for C and N. Thus, it seems that all three elements demonstrate similar patterns of behavior among low-metallicity stars, although variations in O have not been systematically studied in the same manner as for C because of difficulties with detecting O spectral features at low metallicity. Some N measurements are available, but are generally also very difficult to obtain in low-metallicity stars.
We also find a similar behavior for Na with 
a weak but still statistically significant correlation coefficient of 0.22. 
Furthermore, when dividing the data into sodium-rich ([Na/Fe] $>$ 0.7) and regular stars, the correlation between Fe and Na increases to 0.7 for regular stars, in a similar way as for C, N, and O. 

Overall, although CEMP, NEMP as well as O and Na enhanced stars are not identical in their behavior, their progenitor stars were likely faint (black-hole forming) supernovae that experienced substantial mixing (of matter from the inner zones during the explosion mechanism before falling into the nascent black hole) with the outer ejected zones that stem from earlier hydrostatic burning stages. Small Fe ejecta of these first zero-metallicity stars lead to a weak correlation with C, N, O, and Na, of order 0.2. In contrast, the regular EMP stars, not very enhanced in C, N, O, or Na, show a strong correlation around 0.8 and point to regular CCSNe as their progenitors.

\section{CEMP-s, CEMP-rs, and CEMP-r stars}

In the previous Section on correlations of C, N, O, and Na with Fe we discussed the behavior (and probable origin) of CEMP stars. Here we will consider certain abundance patterns of heavy elements which accompany the CEMP star abundance patterns of light and intermediate mass elements. Observers have introduced the categories CEMP-s, CEMP-rs, and CEMP-r stars, which describe carbon-enhanced stars that also display significant overabundances of heavy s-process elements that follow a characteristic pattern, a heavy-element pattern that matches neither the s-process nor the r-process pattern (but in some cases that of an intermediate i-process; \citealt{Hampel.ea:2016}), and a pure main r-process pattern of heavy elements, respectively. 

Generally, galactic chemical evolution modeling suggests that significant s-process enrichment of the interstellar medium only sets in at metallicities of about [Fe/H] $= -$2.5 to -2 due to the delayed onset of low- and intermediate mass stars that are undergoing the asymptotic giant branch stage when the s-process occurs. Hence, we assume that at metallicities of [Fe/H]$<-$2, the s-process is not a significant source of heavy elements. 

However, some individual stars, down to [Fe/H] $\sim-$2.8, show substantial enhancements in s-process elements along with large amounts of carbon, i.e. CEMP-s stars. These s-process elements do not originate from the birth gas cloud, but are the result of a mass-transfer event. For stars in a binary system, the stellar surface of the CEMP-s star was polluted by the AGB winds of a slightly more massive, more evolved companion star that underwent s-process nucleosynthesis. The binary nature of many CEMP-s stars has been shown, which supports this scenario
\citep{Abate.Pols.Stancliffe:2018}. \cite{Hansen.Andersen.ea:2016} found that out of 22 observed CEMP-s stars, for 18 of them a binary companion was identified. The question remains whether for some (apparently) single CEMP-s stars also the early pollution of the protostellar cloud by a progenitor supernova was possible. \cite{Choplin.Hirschi.ea:2017} could show that at least for three of the four single CEMP-s stars a contribution of a primary s-process from fast rotating spin stars was possible. 

The origins of CEMP-rs stars are more ambiguous as these observed heavy-element abundances show variations and do not follow one characteristic pattern. A significant group of stars stars have been explained \citep{Hampel.ea:2019,Mashonkina.ea:2023} as having formed from gas enriched by progenitors undergoing an i-process, i.e. essentially a very strong s-process which also produces nuclei off stability and therefore resembles in the abundance pattern an s- and r-superposition. Only one star has thus far been found to exhibit a pattern that can be described as a combination of both and s- and r-patterns \citep{Gull.Frebel.ea:2018,Koch.Reichert.ea:2019}, i.e., a star that formed from r-process rich gas and then received a mass transfer event of s-process elements.

\begin{figure}
    \centering
        \includegraphics[width=0.75\linewidth]{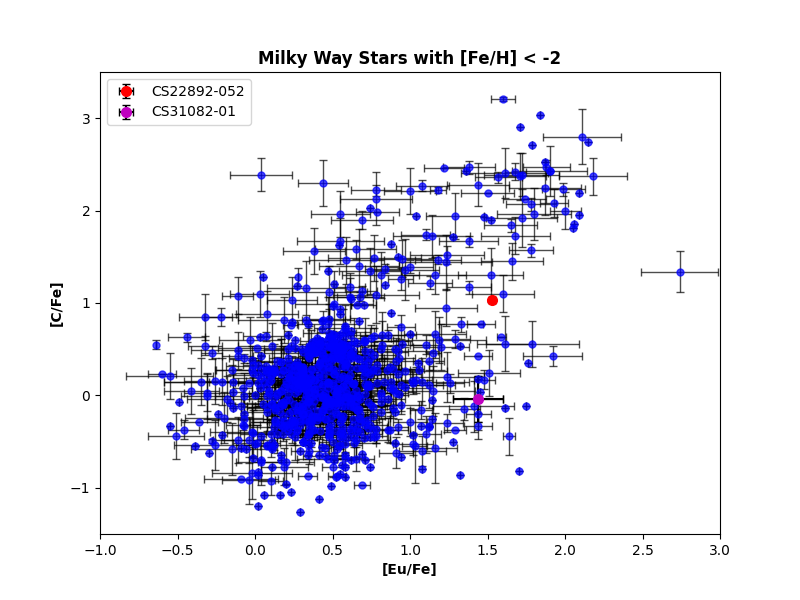}
    \includegraphics[width=0.75\linewidth]{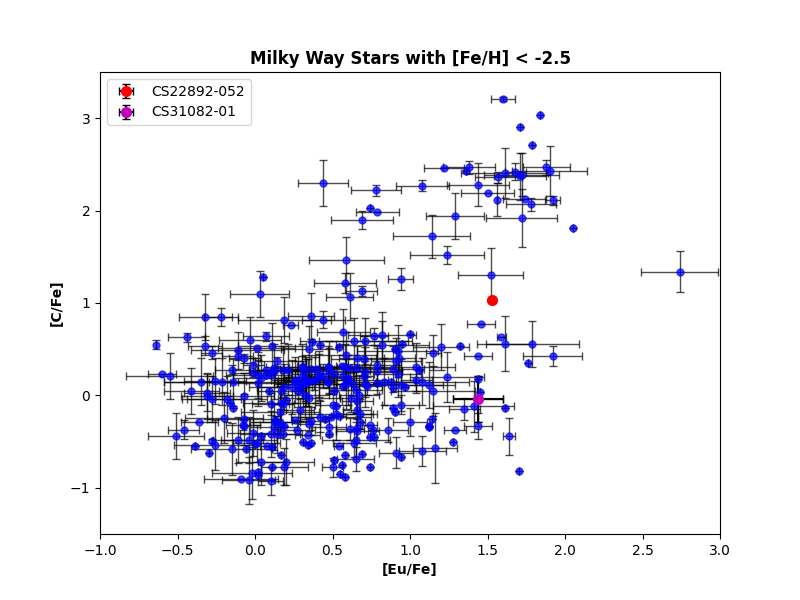}
          \includegraphics[width=0.75\linewidth]{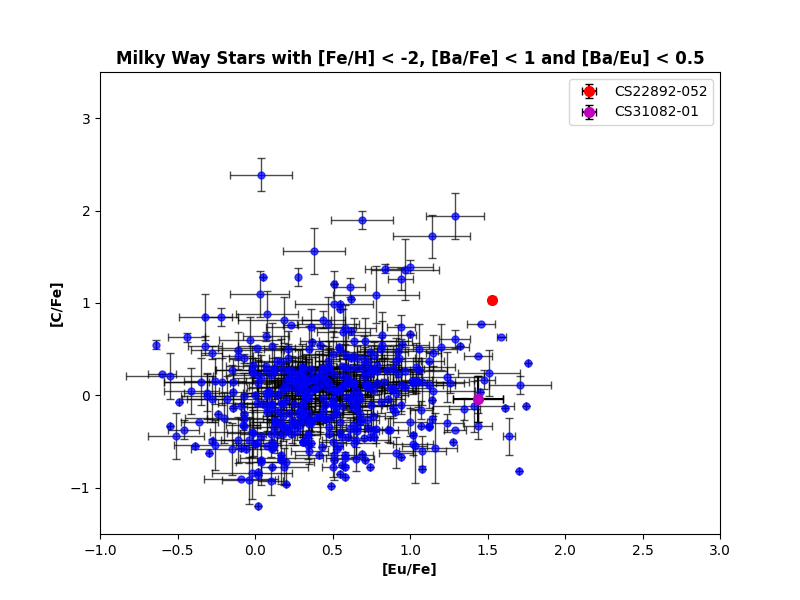}
    \caption{[C/Fe] vs. [Eu/Fe] abundances in VMP/EMP stars with [Fe/H]$<-$2 from the SAGA database. Top: all Milky Way stars with with C, Fe, and Eu measurements for [Fe/H]$<-2$, middle: same as top but only stars with [Fe/H]$<-2.5$, bottom: stars with [Fe/H]$<-2$ as in top figure, but without any stars showing an s-process signature which is always accompanied by large C abundances. Indicated are two example r-II stars by \cite{sneden03} and \cite{cayrel01}, one showing no C-enhancement but an actinide boost, the other one being also a CEMP star.}
    \label{fig:CEMPr}
\end{figure}

Fig.~\ref{fig:CEMPr}, shows the carbon abundances [C/Fe] for regular and CEMP stars with [C/Fe]$>$1 (including various sub-categories) as a function of [Eu/Fe] which represents any r-process contribution. Across panels, we do not distinguish between different categories such CEMP stars and regular low-metallicity stars, or stars that show mild to strong enhancement in r-process elements, from limited-r stars with [Eu/Fe]$<$0.3, to r-I stars with 0.3$<$[Eu/Fe]$<$0.7, and r-II stars with [Eu/Fe]$>$0.7 \citep{Holmbeck.Hansen.ea:2020}, including the two stars  with [Eu/Fe]$>$2 \citep{Cain.ea:2018}. 
The top panel shows all stars with [Fe/H]$<-$2 without further restrictions. This includes stars such as CEMP-s and CEMP-rs stars that are thought to have received their large carbon abundances from binary AGB star companions. These stars typically have [Fe/H]$>-$2.5 
In order to reduce the contamination of our sample by these externally enriched CEMP stars, in the middle panel, we show the same sample but restricted to a metallicity of [Fe/H]$<-$2.5. This removes many of these stars. In the bottom panel, a further restriction of [Ba/Fe]$<$1 and [Ba/Eu]$<$0.5 was applied which strictly excludes any stars with an s-process pattern. This shows that the middle panel still includes some CEMP-rs stars which can have metallicities lower than [Fe/H]$<-$2.5.

Our final CEMP sample thus reveals a small number of r-I and r-II stars are also CEMP stars.
The limited-r stars with [Eu/Fe]$<$0.2 to 0.3, characterized by enhancements in light r-process elements only and with progressively lower abundances for heavier elements,
seem to show a CEMP signature less frequently.
As a caveat, the question remains whether for both the CEMP as well as the r-process enhanced stars with [Fe/H] $=-$2.5 to $-$2, the observed abundances of these elements are due to a combination of the yields of multiple regular CCSNe, one CCSN plus a (later) r-process event but that still occurred prior to the birth of the observed stars, or rather just one single enrichment event (CCSN or r-process event).


There exist many stars for which the observed abundances can be explained in the way that just one regular CCSN contributed to the birth gas cloud \citep{Nomoto.Tominaga.ea:2006,frebel:10,keller:14,Yong.ea:2021}. For example, in the case of the CEMP star SMSS 0313$-$6707 with [Fe/H] $<-7.0$ \citep{keller:14}, that one progenitor appears to have been a faint SN. However, for the EMP star SMSS J200322.54$-$114203.3 \citep{Yong.ea:2021} a pre-enrichment scenario involving a hypenova/collapsar is favored.
On the other hand \cite{Placco.ea:2023} and \cite{Mishenina.ea:2024} find abundance patterns which can be explained by the combination of a regular CCSN and an additional strong r-process source. \cite{cayrel01} found an r-II star, CS 31082-001, (with actinide boost) that is not a CEMP star, while \cite{sneden03} discovered an r-II star, CS22892-052, which is a CEMP star. 

Overall, a wide range of r-process enhancements is found among low-metallicity stars, from limited-r stars to r-I stars and r-II stars, some with carbon enhancement but most without. Hence, the question remains whether these abundance patterns originate from a single enrichment event of the pre-stellar cloud, or whether there are multiple sources and sites involved. In the next Section, we will analyze the metallicity evolution of these r-process stars, i.e. for limited-r, r-I, and r-II stars, following \citet{Farouqi.Thielemann.ea:2022}, considering also to what extent elements up to the Fe-group and r-process elements are co-produced.


In summary, all CEMP stars have small (but significant) correlations of C, N, O and Na with Fe, while the heavy elements show no correlation with Fe. This is easily understandable for CEMP-s and CEMP-rs stars whose abundance patterns can be explained with the completely independent surface pollution mechanism at late times by a binary companion. For CEMP-r stars with r-I and r-II enrichment levels no correlation of the heavy elements with Fe is found. Combining the correlation behavior of C-Na with that of heavy elements thus argues for a independent r-process pre-enrichment scenario. We note that if the [r/Fe] ratio in collapsars/hypernovae is very large, Fe would not be statistically detectable in our correlation analysis. The same outcome would be found if the second source would be a compact binary merger without any Fe production.

\section{Metalliciy Dependence of the Onset of Abundance Features}
\label{onset}

In the previous Sections, we considered the correlations of the discussed elements with Fe over the entire low metallicity range of [Fe/H] $<-2$. In an earlier investigation, \citet{Farouqi.Thielemann.ea:2022} noticed that among stars with heavy elements, the limited-r, r-I, and r-II stars seemed to appear first at metallicity thresholds of about [Fe/H] $\approx$ $-$4, $-$3.7, and $-$3.4 (see also Fig.~\ref{fig:EuHEuFe}). It remains unclear to what extent this sequence maps onto the different origins of these elements from magneto-rotational supernovae, collapsars/hypernovae and possibly compact binary mergers. Hence, it is important to investigate further the effects of the single and multiple origins of these elements. 

The Pearson correlations detect solely linear correlations, i.e. if element X would increase with Fe, but on a quadratic scale, Pearson would not detect it. On the other hand, the Spearman correlation looks only at an increase of element X with the increase of Fe, and therefore can identify a different or additional production site of element X with respect to that of Fe. Considering e.g. Eu, for stars with low abundance ratios of [Eu/Fe] $<$ 0.3, both these correlations find identical behaviors with respect to Fe. However, for stars with larger [Eu/Fe] ratios, a deviation sets in that points to the appearance of a further production site with a weak linear correlation. Next, after removing the limited-r stars from the entire sample and carrying out the same analysis, the combined r-I and r-II star-only sample also starts to show a deviation at a certain [Eu/Fe] ratio.

\begin{figure}
    \centering
     \includegraphics[width=0.8\linewidth]{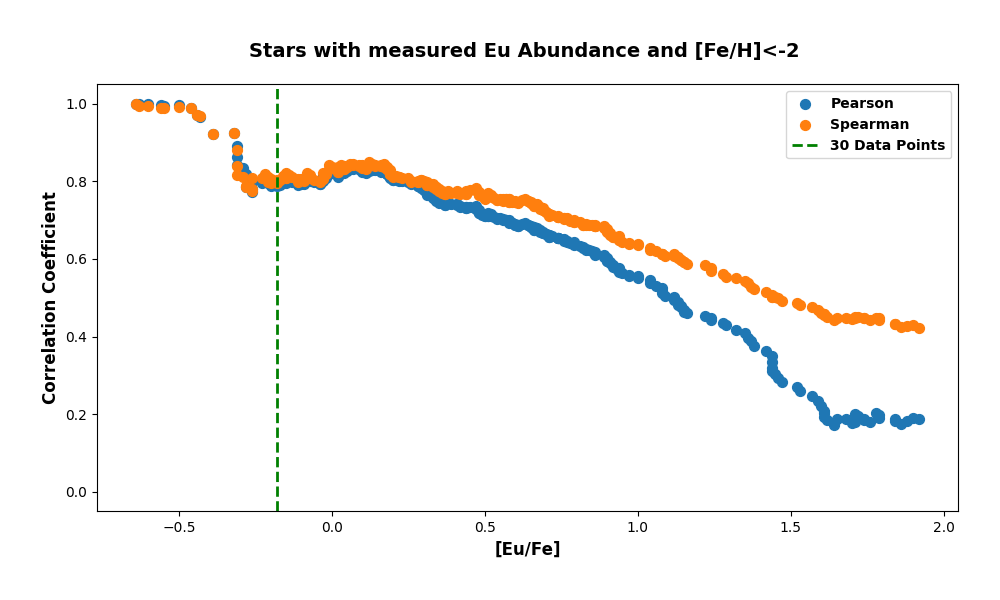}
     \includegraphics[width=0.8\linewidth]{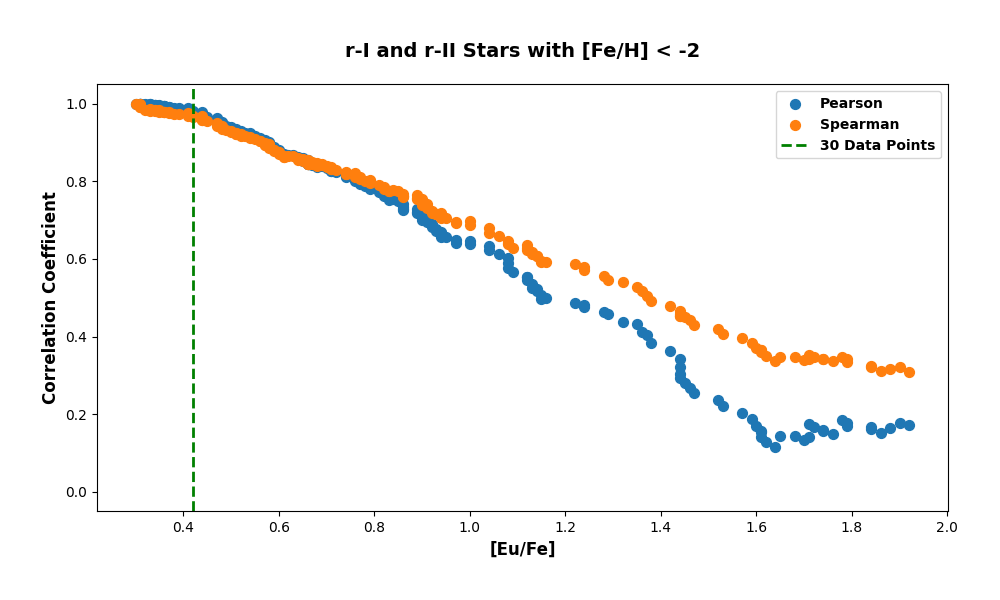}
    \caption{Pearson and Spearman correlation coefficients for Fe and Eu. Top: When utilizing the entire sample of heavy element containing stars (limited r-process, r-I and r-II stars), both correlations lead to identical values for [Eu/Fe] $<$ 0.3, i.e. up to the limit of limited-r stars, while a deviation shows up for larger values, corresponding to r-I and r-II stars. This indicates that a different astrophysical source is required for these two types of heavy element containing stars (limited-r stars vs r-process enhanced stars). Bottom: When excluding limited-r stars from the sample, a deviation appears at [Eu/Fe] = 0.7, i.e. at the onset of r-II stars, also pointing to the need of a additional independent source for these elements. [ the 'all stars' label at the top graph should be specified to 'limited r-process stars, rI- and r-II stars]}
    \label{fig:PCCSp}
\end{figure}

This behavior is shown in Fig.~\ref{fig:PCCSp}. When making use of all stars with [Fe/H] $<-2.0$ in the SAGA database, we see that at [Eu/Fe] = 0.3 the Pearson and Spearman correlation coefficients start to deviate from each other, while they have identical values up to that point. Above this [Eu/Fe] value, the relation between Fe and Eu is not linear anymore .which we interpret as a contribution of a different source of these elements. When only taking the sample of r-I and r-II stars (i.e. excluding limited-r stars), a similar behavior sets in at [Eu/Fe] = 0.7, i.e. just at the transition between the r-I and r-II stars. Again, this indicates the appearance of an additional and different source that would be responsible for the abundance patterns of r-II stars. Thus, our analysis of the observed abundance patterns suggests that (at least) three astrophysical sources and/or sites are required to explain the existence of the limited-r, r-I, and r-II stars. It remains to be seen whether the actinide-boost stars would require yet an additional source as well.

\begin{figure}
    \centering
    \includegraphics[width=0.9\linewidth]{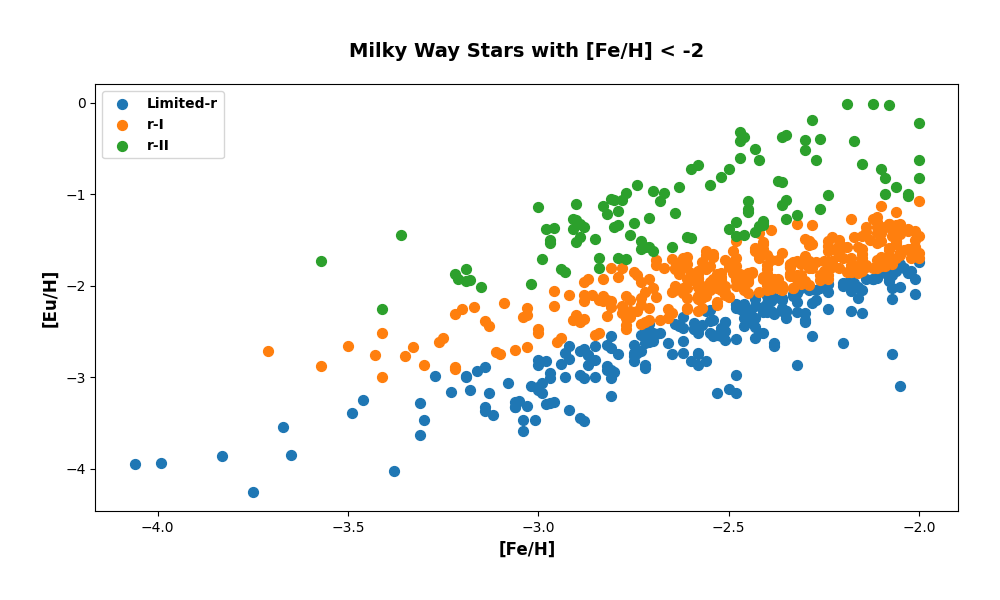}
    \caption{[Eu/Fe] abundances as a function of [Fe/H] for low metallicity limited-r (blue), r-I (orange), and r-II (green) stars with [Fe/H] $<-2$. An apparent
    change in the onset of the three components can be noticed at different metallicities.}
    \label{fig:EuHEuFe}
\end{figure}

One can now ask the question whether these different categories of stars with heavy elements associated with the r-process emerged at different times during the galactic chemical evolution. This would be reflected in their metallicity distributions.   
Fig.~\ref{fig:EuHEuFe} displays stars with available Eu and Fe] abundances in the [Eu/Fe] vs. [Fe/H] plane. The onset of the three categories indicates clearly that the limited-r pattern appears in the most metal-poor (earliest?) EMP stars for metallicities below [Fe/H] $\approx-$4. The r-I and r-II stars follow at [Fe/H] $\approx-$3.5. 

One could then compare this behavior with the expected frequency of events. If we associate limited-r stars with the contribution from magneto-rotational supernovae, i.e. supernovae that produce fast spinning pulsars with high magnetic fields (magnetars), this amounts to about 10\% of regular CCSNe \citep[e.g.][]{Beniamini.Hotokezaka.Horst.ea:2019}.
One expects the lowest-metallicity stars to have formed from gas enriched by the first CCSNe that had yields commensurate with about [Fe/H] $\approx-$5, or even less \citep[e.g.][]{Norris.Christlieb.ea:2007,Norris.Christlieb.ea:2012,Frebel.Norris:2015,Nordlander.ea:2017}. For events which occur with 1/10 of the frequency of regular CCSNe (like magneto-rotational supernovae leading to magnetars) one might then expect them to emerge first at metallicities around [Fe/H] $\approx-$4, consistent with the findings for limited-r stars in Fig.~\ref{fig:EuHEuFe}. 

Nucleosynthesis events that take place only after about 100 regular CCSNe enriched the ISM, can be expected to lead to stars with  [Fe/H]$\approx$-3 and above (\cite{Farouqi.Thielemann.ea:2022} argue that collapsars as well as compact binary mergers have a frequency lower by about a factor of 120 than regular CCSNe). These orders of magnitude estimates appear be reflected in what is shown in Fig.~\ref{fig:EuHEuFe} for the different r-process enhanced stars. While from Fig.~\ref{fig:EuHEuFe} alone one cannot argue that the three categories limited-r, r-I, and r-II stars represent the pre-enrichment scenarios by different sites, we note in concluding that Fig.~\ref{fig:PCCSp} would, however, support this conclusion.

\section{Conclusions}
After having discussed our statistical analysis of element correlations in extremely low-metallicity stars, we once again consider our findings as summarized in the color-coded Table~\ref{tab:correlation_resultsB}. 

Blue colors stand for {\it no correlation at all with Fe (PCC$<$0.1)}. Li stemming only from the Big Bang at these low metallicities is a clear case with no CCSN origin. The heaviest elements from a strong r-process, point also to no CCSN origin and the need that the abundances of these elements have to be explained by other sources. These could be compact binary mergers and/or collapsars which co-produce no Fe (ejected in CCSNe) or with [r/Fe] ratios which are strongly supersolar, causing a vanishing correlation with Fe.

Green stands for a {\it strong correlation with Fe (PCC$>$0.7-0.8)}. This applies to (a) C,N,O from
regular (not CEMP) stars,  (b) $\alpha$-elements up to Ti, (c) all Fe-group elements, plus elements up to Zn and even Se. 
This points directly to a massive star origin (CCSNe) including $\alpha$-elements, the Fe-group as well as also $\nu$p-process abundances from the innermost ejected zones of CCSNe.

Teal stands for a {\it moderately strong correlation with Fe (PCC between about 0.4 and 0.7)}. Somewhat surprisingly, Be appears to be in this group. In the first massive stars (that lack any CNO seed nuclei) Be is produced in H-burning via hot pp-chains and $^{3}$He($\alpha$,p) and later ejected jointly with explosively processed matter. 
The apparent PCC values of Rh, Pd, and Ag among these elements (which would argue for a very weak/limited r-process operating even in regular supernovae) suffer from a small number of data points and as such, the theoretical speculations need to be confirmed by further observations as well as more supernova modeling efforts.

Red color coding stands for a {\it weak, but significant correlations of PCCs in the range 0.1-0.4}. This points to a moderately weak r-process contribution from magneto-rotational supernovae which (a) have a somewhat reduced Fe production in comparison to regular CCSNe and (b) a weak/limited r-process which leaks beyond the second r-process peak and can produce Eu and further rare earth elements. The relatively low PCC level indicates also that important additional r-process source(s) are required in order to explain these abundances.

A major finding of this study is that the Eu and third r-process peak element production are decoupled to a certain extent, due to the fact that a prominent Eu provider co-produces Fe, while the heavy third r-process peak elements are produced independently without Fe. This result agrees with recent findings by \cite{APuls.ea:2024}.

We have also included C, N, O, and Na of CEMP, NEMP, OEMP (and possible NaEMP) stars in this correlation bracket (red colors). Their weak, but still significant, correlations point also to a core-collapse supernova origin of the failed/faint type, which produces Fe, but in very small amounts due to fallback onto the emerging black hole. A much higher correlations for these elements is found in regular CCSNe (color coded in green).

The remaining elements in the Table are given in black. For these, the data are too scarce to argue for a significant correlation. 

We note that, interestingly, we do not find a strong correlation with the second s-process peak element Ba. While a strong correlation with Fe would not be expected for EMP stars because (major) s-process contributions from low and intermediate mass stars are only expected above [Fe/H]$=$-2.5 to -2. 

However, if a strong correlation would have been found, this would have pointed to the s-process operating in massive spin stars \citep{Frischknecht.Hirschi.ea:2012,Frischknecht.Hirschi.ea:2016} where s-process elements are made from primary N and $^{22}$Ne and via shear mixing in the outer layers of these massive stars. S-process elements are then thought to be jointly ejected with the explosively processed supernova matter. But our correlation analysis does not support a significant contribution from such a scenario.

The present review focused mainly on correlations among all the different elements, in particular with respect to Fe. We intended to investigate (various types of) core-collapse supernova origins of these elements as observed in EMP stars. It could also help to determine for which elements additional or even exclusively different types of nucleosynthsis events are needed to explain the observed abundance patterns.




Our overall conclusion, based on the elements colored in green in Table {\ref{tab:correlation_resultsB}}, is that these elements clearly point to their origins from CCSNe. The elements in red point to at least a partial origin from (possibly a specific type of) CCSNe, while they require also an additional origin. Based on Fig.~\ref{fig:EuHEuFe}, we  suggest that after a floor of (regular) CCSN contributions, starting at metallicities as low as [Fe/H] = $-$5 to even $-$6, any subsequent contributions at metallicities from [Fe/H] = $-$4 to $-$3.5 are from r-process events. This could include rare, early magneto-rotational supernovae (occurring at a fraction of about 10\% or all CCSNe) that produce a limited-r star abundance pattern. Further star samples, such as r-I and r-II stars, might point to collapsars and/or compact binary mergers. In the Section on CEMP-r stars, we discussed also how possible superpositions are required, as there exist CEMP stars with and without r-I or r-II enhancements. The onset of different events could point to their associated occurrence frequencies. For further insights into the subject, we refer the interested reader to the comprehensive review by \cite{Frebel:2024}, addressing various observational signatures of low-metallicity stars in an attempt to explain the origin of the many observed abundance patterns among low-metallicity stars. 

\vfil\eject

\backmatter




\bmhead{{\bf APPENDICES}}
In the present paper we applied frequently a range of statistical tools like Pearson and Spearman correlation coefficients, p-values, significance, and mutual information. None
of these concepts are our original work, and for more information we refer to statistics text books like \citep{tamhane00,spiegelhalter19}. In addition, in our precursor paper, dealing only with r-process elements and their production sites, we have already introduced our use of Pearson and Spearman correlation coefficients. For this reason, we do not want to repeat this here. But, opposite to our precursor paper, we concentrated here on testing the significance of the obtained correlation results and utilized mutual information (MI), in order make sure how much one can rely on the obtained conclusions. For this reason we add here appendices on p-values and mutual information.
\begin{appendices}

\section{P-Value and Null Hypothesis}

The concepts of the \textit{p-value} and the \textit{null hypothesis} are fundamental in statistical hypothesis testing, which is widely used in scientific research to evaluate evidence and make data-driven decisions.

The null hypothesis (\(H_0\)) represents the default assumption that there is no effect, no difference, or no relationship between variables in a given study. It serves as the baseline or "status quo". For example, when testing whether a new drug is effective, the null hypothesis might state that the drug has no effect compared to a placebo.

The \textit{p-value}, or \textit{probability value}, quantifies the evidence against the null hypothesis. It represents the probability of observing results as extreme as (or more extreme than) those obtained in the study, assuming that the null hypothesis is true.

The purposes of these two concepts are:
\begin{itemize}
    \item The \textbf{null hypothesis} provides a framework for statistical testing, serving as the hypothesis to be tested against an alternative hypothesis (\(H_1\)), which proposes a specific effect or relationship.
    \item The \textbf{p-value} helps researchers decide whether to reject the null hypothesis. A smaller p-value indicates stronger evidence against the null hypothesis, while a larger p-value suggests insufficient evidence to do so.
\end{itemize}

The correct interpretation of these concepts is:
\begin{itemize}
    \item \textbf{Small P-Value (e.g., \(p < 0.05\))}:  
    The observed data are unlikely under the null hypothesis. This typically leads to rejecting the null hypothesis in favor of the alternative hypothesis.
    \item \textbf{Large P-Value (e.g., \(p \geq 0.05\))}:  
    The observed data are consistent with the null hypothesis. This indicates insufficient evidence to reject the null hypothesis.
\end{itemize}

Among other things, the two concepts are used for:
\begin{enumerate}
    \item \textbf{Decision-Making in Research:}  
    P-values are used to determine if experimental results are statistically significant, such as evaluating the effectiveness of treatments or interventions.
    \item \textbf{Setting a Significance Level (\(\alpha\)):}  
    Before conducting a test, researchers set a significance threshold (commonly \(\alpha = 0.05\)). If the p-value is less than \(\alpha\), the null hypothesis is rejected.
    \item \textbf{Reporting Results:}  
    P-values are often reported in research papers to quantify the strength of evidence against the null hypothesis.
\end{enumerate}

The most well-known limitations are:
\begin{itemize}
    \item \textbf{Does Not Prove Truth:}  
    A p-value does not prove that the null hypothesis is true or false; it merely quantifies the strength of evidence against it.
    \item \textbf{Context Matters:}  
    Statistical significance does not always imply practical or scientific importance. Other factors, such as effect size and study design, should also be considered.
    \item \textbf{Risk of Misinterpretation:}  
    P-values are often misinterpreted as the probability that the null hypothesis is true, which is incorrect.
\end{itemize}

The \textit{p-value} and \textit{null hypothesis} are essential tools for understanding data, but they require careful interpretation to avoid misleading conclusions. They are most effective when used in combination with other statistical measures and domain knowledge.

\section{Mutual Information}

Mutual Information (MI) is a measure of the mutual dependence between two random variables. It quantifies the amount of information obtained about one variable through the knowledge of the other variable. Formally, the Mutual Information between two variables \(X\) and \(Y\) is defined as:

\begin{equation}
I(X; Y) = \sum_{y \in Y} \sum_{x \in X} P(x, y) \log \left( \frac{P(x, y)}{P(x) P(y)} \right),
\end{equation}

where \(P(x, y)\) is the joint probability distribution of \(X\) and \(Y\), and \(P(x)\) and \(P(y)\) are the marginal distributions of \(X\) and \(Y\), respectively.

Mutual Information has several useful properties:
\begin{itemize}
    \item \textbf{Non-negativity}: MI is always greater than or equal to zero (\(I(X; Y) \geq 0\)). \(I(X; Y) = 0\) means that the variables \(X\) and \(Y\) are independent.
    \item \textbf{Symmetry}: Mutual Information is symmetric, meaning \(I(X; Y) = I(Y; X)\).
    \item \textbf{Information Gain}: MI can be viewed as the expected reduction in uncertainty about \(X\) when \(Y\) is known, and vice versa.
\end{itemize}
\textbf{Applications of Mutual Information in Research and Data Science}\\
Mutual Information finds wide application in various fields of research and data science:

\begin{itemize}
    \item \textbf{Feature Selection}: In feature selection, MI-based methods are used to identify relevant features from large datasets. MI helps in selecting those features that provide the most information about the target variable.
    
    \item \textbf{Cluster Analysis}: MI can be used to assess the quality of cluster assignments. It quantifies how well the generated clusters capture the underlying data patterns.
    
    \item \textbf{Image and Speech Processing}: In image and speech processing, MI is often used for registration and alignment of images. The goal is to maximize the mutual information between two images to achieve accurate matching.
    
    \item \textbf{Dependency Analysis}: MI is employed to quantify dependencies and relationships between variables in a dataset. This is particularly useful in fields like genomics, where understanding the interactions between genes is crucial.
    
    \item \textbf{Information Gain in Decision Trees}: In machine learning algorithms such as decision trees, MI is used to guide the most informative splits by maximizing the information gain.
\end{itemize}

Mutual Information is a powerful tool for analyzing relationships between variables and is widely used in various applications in research and data science to gain insights and improve models. Its ability to capture non-linear relationships makes it particularly valuable in complex data environments.

\end{appendices}

\bmhead{Acknowledgements}

We want to thank our colleagues John Cowan, Tamara Mishenina, and Ian Roederer for very fruitful discussions on available observational data.

The current research benefitted from the COST Action CA 16117 (ChETEC) and the US National Science Foundation under grant  PHY-1927130 (AccelNet-WOU: International Research Network for Nuclear Astrophysics [IReNA]).
A.F. also acknowledges support from the US National Science Foundation under grants AST-2307436 and PHY-1430152 (JINA Center for the Evolution of the Elements)


\end{document}